\documentclass[aoas]{imsart}

\RequirePackage{amsthm,amsmath,amsfonts,amssymb}
\RequirePackage[authoryear]{natbib}
\RequirePackage[colorlinks,citecolor=blue,urlcolor=blue]{hyperref}
\RequirePackage{graphicx}
\RequirePackage{threeparttable}
\RequirePackage{multirow}
\RequirePackage{algorithm}
\RequirePackage{algorithmic}

\startlocaldefs
\theoremstyle{plain}

\newtheorem{theorem}{Theorem}

\theoremstyle{remark}

\endlocaldefs

\begin{document}

\begin{frontmatter}
\title{Identification of Influencing Factors on Self-reported Count Data with Multiple Potential Inflated Values}
\runtitle{MINB}

\begin{aug}
\author[A,B]{\fnms{Yang}~\snm{Li}\ead[label=e1]{yang.li@ruc.edu.cn}},
\author[A]{\fnms{Mingcong}~\snm{Wu}\ead[label=e2]{wumingcong@ruc.edu.cn}},
\author[C]{\fnms{Mengyun}~\snm{Wu} \thanksref{t1}\ead[label=e3]{wu.mengyun@mail.shufe.edu.cn}},
\and
\author[D]{\fnms{Shuangge} \snm{Ma}\ead[label=e4]{shuangge.ma@yale.edu}}
\address[A]{Center for Applied Statistics and School of Statistics, 
Renmin University of China, Beijing, China\printead[presep={,\ }]{e1,e2}}

\address[B]{Institute for Data Science in Health, Renmin University of China, Beijing, China}

\address[C]{School of Statistics and Management, 
Shanghai University of Finance and Economics, Shanghai, China\printead[presep={,\ }]{e3}}

\address[D]{Department of Biostatistics, 
Yale School of Public Health, New Haven, USA\printead[presep={,\ }]{e4}}

\thankstext{t1}{corresponding author: wu.mengyun@mail.shufe.edu.cn}

\end{aug}

\begin{abstract}
The Online Chauffeured Service Demand (OCSD) research is an exploratory market study of designated driver services in China. Researchers are interested in the influencing factors of chauffeured service adoption and usage and have collected relevant data using a self-reported questionnaire. As self-reported count measure data is typically inflated, there exist challenges to its validity, which may bias estimation and increase error in empirical research. Motivated by the analysis of self-reported data with multiple inflated values, we propose a novel approach to simultaneously achieve data-driven inflated value selection and identification of important influencing factors. In particular, the regularization technique is applied to the mixing proportions of inflated values and the regression parameters to obtain shrinkage estimates. We analyze the OCSD data with the proposed approach, deriving insights into the determinants impacting service demand. The proper interpretations and implications contribute to service promotion and related policy optimization. Extensive simulation studies and consistent asymptotic properties further establish the effectiveness of the proposed approach.

\end{abstract}

\begin{keyword}
\kwd{inflated value selection}
\kwd{investigation}
\kwd{negative binomial data modeling}
\kwd{penalization}
\end{keyword}

\end{frontmatter}

\section{Introduction}
\label{sec:intro}
In this era of data deluge, the emergence and progress of network technology have advanced network services, bringing new opportunities and challenges to the development and management of the online service industry. This study is motivated by an Online Chauffeured Service Demand (OCSD) research conducted by the world's leading mobile transportation platform, which offers application-based mobility services. The platform gained revenue of over RMB 130 billion in 2020 from its domestic mobility business. Designated driving, as one of its core businesses, has acquired a rising number of orders at a high rate for consecutive years since its launch in 2015. The number of chauffeur drivers who earned income through the platform in 2019 was 229,000, reporting a year-on-year growth of up to 27\%. 

\begin{figure}[htbp]
\centering
\includegraphics[width=0.9\textwidth]{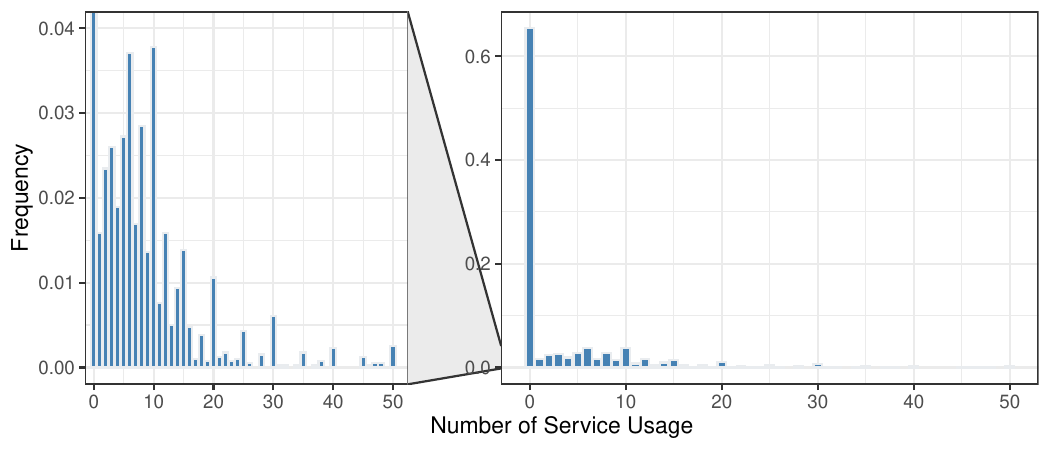}
\caption{Frequency histogram of the observed self-reported counts. Left: zoomed-in version with the x-axis from 1 to 50 and the y-axis from 0.00 to 0.04.} \label{freq-original}
\end{figure}

The rapid growth in demand and market size for chauffeur services has created new development opportunities for the platform. To better understand service demand and address market changes, the platform conducted the OCSD study to attain data support on marketing assessment and potential customer management. This investigation targeted to generate influencing factor analysis on designated driving service demand, advancing insights for personalized customer management and strategic marketing. The target response was defined as the frequency of chauffeured service usage during the past six months. Figure \ref{freq-original} presents the observed self-reported counts with noticeable inflated integers, indicating that the observed responses do not follow a ``standard'' count distribution, such as Poisson or Negative Binomial (NB). Specifically, with the presence of potential users who have not adopted the service, the majority of the response usage is zero (about 64 $\%$). Additionally, there are evident inflation at multiples of 10, probably because people tend to report rounded approximations.

The aforementioned inflation can distort the distribution of the observed counts and lead to biased estimation and increased standard error \citep{Zhu2015Variable,cai2021generalized}. This issue has been widely discussed in self-reported count data regression. \cite{wang2008modeling} considered rounded-off cigarette consumption data with inflation on values of 5, 10, and 20, where such reporting error introduced biased estimation of mean cigarette consumption. \cite{bocci2021multiple} examined the tourism behavior of Italians using the multiple-inflated model, which correctly fitted the distribution of the total number of overnight stays, whereas the extremely inflated values were under-predicted by the standard models. To demonstrate the issue more intuitively, we use a toy example to evaluate the impact of inflation on regression parameter estimates. Specifically, we consider a multiple-inflated Poisson distribution and generate covariates from the normal distribution $N(0,1)$ with the true regression parameters $(-2,-5,1)^\top$. We assume inflation on 0, 1, 10, and 20 with the inflated proportions ranging from $0\%$ to $20 \%$. The Poisson regression is fitted on the inflated observations, and Table \ref{tab1} provides the mean values of parameter estimates based on 100 replications. It is observed that the influence on estimates increases as the inflation rate rises. Specifically, when $80 \%$ of the data is inflated, the estimates are 0.328 for a true value of -2 and -0.002 for a true value of 1. Similarly, in the analysis of the OCSD data, without an accommodation for multiple inflated values, classic methods such as Poisson and NB regression may provide unreasonable findings that ``severe government punishment'' on drunk driving has a negative effect on chauffeured service demand (we refer to Section 4 for more details).
 
\begin{table}[htbp]
\centering
\renewcommand\arraystretch{1}
\setlength\tabcolsep{12pt}
\caption{Estimates of regression parameters with Poisson regression for multiple-inflated Poisson observations. In each cell, the mean value is based on 100 replicates.}
\begin{tabular}{ccr@{.}lr@{.}lr@{.}lr@{.}l}
\hline
      &            & \multicolumn{4}{c}{Inflation rate} \\
      & True value & \multicolumn{2}{c}{0\%}    & \multicolumn{2}{c}{20\%}       & \multicolumn{2}{c}{40\%}      & \multicolumn{2}{c}{80\%}      \\ \hline
$\beta_1$ & -2         & -2&015    & -0&300    & -0&138    &  0&328    \\
$\beta_2$  & -5         & -4&981    & -0&667    & -0&270    & -0&073   \\
$\beta_3$  & 1          & 1&017     & 0&171     & 0&053     & -0&002   \\ \hline
\end{tabular}
\label{tab1}
\end{table}

Numerous methods have been developed to
address the problem of inflation. The zero-inflated Poisson (ZIP) model was first proposed and applied in the data analysis of manufacturing defects \citep{lambert1992zero}. Then, the zero-inflated NB (ZINB) model \citep{garay2011on} was introduced to further account for overdispersion detected in an apple cultivar data set. Subsequently, \cite{xie2014bayesian} proposed a zero-inflated generalized Poisson (ZIGP) distribution model as a useful generalization of ZIP. \cite{buu2011new} and \cite{Zhu2015Variable} proposed penalized variable selection methods for the ZIP and ZINB models, respectively. \cite{ZIGDM} developed a novel probability distribution, the zero-inflated generalized Dirichlet multinomial (ZIGDM), to flexibly accommodate complex correlation structures and dispersion patterns.  \cite{giles2007modeling} generalized the ZIP model to a multiple-inflated Poisson model (MIP), enabling counting inflation at multiple points. Furthermore, \cite{giles2010hermite} introduced the Hermite distribution with the capability of modeling multi-inflated count data and accounting for overdispersion. \cite{wang2008modeling} presented a Bayesian model to describe heaped count data from a randomized trial of smoking cessation. \cite{2019li} proposed a two-stage inflated value selection method under the framework of MIP. \cite{cai2021generalized} proposed a generalized inflated discrete model (GIDM) for various types of discrete outcomes with multiple inflated values. \cite{yee2022generally} developed a Generally, Altered, Inflated, Truncated, and Deflated Negative Binomial (GAITD-NB) model to jointly describe potential alteration, inflation, truncation, and deflation patterns in count data.
 
The aforementioned and other studies have mostly prioritized the zero-inflation situation, with relatively little focus on multiple-inflated data modeling. The MIP model was developed based on the less general Poisson regression, where the variance is equal to the mean. In addition, there is a lack of appropriate accommodation for the selection of inflated values. Specifically, most of the existing models \citep{giles2007modeling,su2013multiple} have adopted a two-stage strategy wherein potential inflated values are selected by first artificially inspecting the histogram. Then, multiple-inflated models with various combinations of inflated values are fitted. The optimal model is selected by criteria such as the Chi-squared goodness-of-fit, Akaike information criterion (AIC), or Bayesian information criterion (BIC), and the inflated values from the best model are subsequently confirmed. However, the traditional histogram inspection technique is usually subjective, with some missed or wrongly selected inflated values, resulting in the inappropriate establishment of the regression model and problems in the subsequent parameter estimation and statistical inference. In addition, the two-stage approach is usually time-consuming because it requires examining numerous combinations of potential inflated values in order to get good results. The selection of inflated values in the first place has still not been well investigated, despite the widespread presence of multiple inflated values in count data analyses.

Motivated by challenges from the analysis of the OCSD data and those alike, we propose a novel inflated value selection approach to handle multiple potential inflated values. We advance the existing count data modeling \citep{giles2007modeling,2019li} by proposing a fully generalized Multiple-Inflated NB (MINB) model that allows for count inflation at a multiplicity of values. This model can effectively accommodate extra overdispersion due to inflation and is more general as NB has fewer constraints than Poisson. The penalization technique is novelly adopted to achieve data-driven inflated value selection and is more effective and computationally efficient than the existing two-stage strategy. Among many potential influencing factors, researchers are often interested in identifying a small subset to offer more lucid interpretations. Thus, we also incorporate variable selection for the purpose of identifying important factors. Statistical properties of the estimator are carefully established, which have not been well investigated for most of the alternatives. The effectiveness of the proposed analysis is also verified by extensive simulation studies. The analysis of the OCSD data leads to statistically satisfactory and commercially meaningful results. The identified important factors are markedly favorable to the usage of chauffeured service, together with those that can become barriers. Moreover, they can contribute to the existing literature regarding the understanding of personalized chauffeured service and provide suggestions for platforms to promote service and make relevant management policies.

\section{Methods}
\label{sec:meth}
\subsection{Penalized MINB Model}

Consider $n$ independent observations. For the $i$th observation, denote $y_i$ as the count response of interest and $\boldsymbol{X} _i =\left( X_{i1},\dots,X_{ip}\right) ^\top$ as the $p$-dimensional covariates. We consider the multiple inflated NB model and assume that the probability mass function of $y_i$ given $\boldsymbol{X}_i$ is :  
\begin{equation}\label{eq:model}
f \left(y_i; \boldsymbol{X}_i, \boldsymbol{\theta} \right) =  \left(  \sum_{j=1}^{J} \omega_{j} \cdot \text{I} \left( y_i = k_j \right) \right) +\omega_{J+1} \cdot  f_{\text{NB}}\left( y_i;  \exp \left(\alpha+ \boldsymbol{X}_i^\top \boldsymbol{\beta} \right) ,\phi \right),
\end{equation}
with $f_{\text{NB}}\left( y_i;  \exp \left(\alpha+ \boldsymbol{X}_i^\top \boldsymbol{\beta} \right) ,\phi \right) =  \frac{\Gamma(y_{i}+\phi^{-1})}{y_{i}!\,\Gamma(\phi^{-1})}\left( \frac{\exp \left(\alpha+ \boldsymbol{X}_i^\top \boldsymbol{\beta} \right) \phi}{1+ \exp \left(\alpha+ \boldsymbol{X}_i^\top \boldsymbol{\beta} \right) \phi}\right) ^{y_{i}}\left( \frac{1}{1+ \exp \left( \alpha+\boldsymbol{X}_i^\top \boldsymbol{\beta} \right)\phi}\right) ^{\phi ^{-1}} $ and $\text{I} (\cdot)$ being the indicator function. Here, $J$ is the number of the multiple inflated values $\boldsymbol{K} = \left \{ k_1, \dots, k_J \right \}$ and $ \boldsymbol{\omega}=\left( \omega_1,\dots,\omega_{J+1}\right) ^\top$ is the vector of mixing proportions satisfying $ \omega_j \geq 0 $ and $\sum_{j=1}^{J+1}\omega_j = 1$. $\phi$,  $\alpha$, and $\boldsymbol{\beta}=\left(\beta_1,\dots,\beta_p\right) ^\top$ are the unknown dispersion parameter, intercept, and coefficient vector, respectively, and $\boldsymbol{\theta} = \left( \phi, \alpha, \boldsymbol{\beta}^\top, \boldsymbol{\omega}^\top \right)^\top \triangleq\left(\theta_j\right)_{ \left( 1+(p+1)+ (J+1) \right)  \times 1} $. 

In practice, the inflated values are not always observed, and not all covariates are associated with the response. Given the candidate inflated values $k_1,\cdots, k_J$, and $p$-dimensional covariates, for estimation and selection of both inflated values and covariates, we propose the penalized objective function: 
\begin{align}
\max \limits_{\boldsymbol{\theta}} pl(\boldsymbol{\theta}) = &l\left(\boldsymbol{\theta}\right) - n \sum_{j = 1}^p p_{\lambda_{1n}} \left( \beta_j  \right) - n \sum_{j = 1}^J p_{\lambda_{2n}} \left( \omega_j \right)  \notag \\
= &\sum_{i=1}^n \log \left \{  \left( \sum_{j=1}^{J}  \omega_{j} \cdot \text{I} \left( y_i = k_j \right) \right)  +\omega_{J+1} \cdot    f_{\text{NB}}\left( y_i;  \exp \left(\alpha+ \boldsymbol{X}_i^\top \boldsymbol{\beta} \right) ,\phi \right) \right \}  \notag \\
&  - n \lambda_{1n} \sum_{j = 1}^p \rho_{1j}  \left | \beta_j  \right| - n \lambda_{2n} \sum_{j=1}^{J} \rho_{2j} \omega_j   \notag, \\
& \quad \quad  \text{subject to } 0\leq \omega_j\leq 1\text{ and }\sum_{j=1}^{J+1}\omega_j = 1,
\label{eq:obj}
\end{align}
where $\boldsymbol{\rho}_1 = \left( \rho_{11}, \dots, \rho_{1p} \right)^{\top}$  and  $ \boldsymbol{\rho}_2 = \left( \rho_{21}, \dots, \rho_{2J} \right)^{\top}$ are the adaptive weights for covariates and inflated values respectively, and $\lambda_{1n}$ and $\lambda_{2n}$ are the corresponding tuning parameters. The proposed estimate is defined as the solution of (\ref{eq:obj}). The nonzero components of $\omega_1,\cdots,\omega_J$ and $\boldsymbol{\beta}$ correspond to the identified inflated values and important covariates, respectively.

In (\ref{eq:obj}), the first term is the log-likelihood, where the mixing proportions $\omega_1,\cdots,\omega_J$ are introduced to accommodate the potential multiple inflated values and the NB distribution is introduced to accommodate the over-dispersion and count nature of the chauffeured service usage data. The second and third terms are the adaptive LASSO penalty imposed on the regression coefficient $\boldsymbol{\beta}$ and mixing proportions $\omega_1,\cdots,\omega_J$, respectively. The adaptive LASSO penalty is adopted as it has satisfactory theoretical and numerical performance \citep{Zou2006,banerjee2018research} and enjoys unbiasedness properties without overshrinking for large coefficients. Significantly advancing from the existing studies which mostly focus on identifying important covariates with penalization only \citep{2014zengvariable,2019xiespatial}, we innovatively conduct the selection of inflated values by applying penalization to the mixing proportions. In practice, all count responses can be regarded as potential inflated points, and some of the mixing proportions are shrunk to zero to achieve the selection of credible inflated values. The proposed strategy enjoys the advantage of determining the inflated values automatically and avoiding subjective selection. Since the intercept term and NB regression component are always included in the model, $\alpha$ and $\omega_{J+1} $ are not subject to penalization. In addition, $\sum_{j=1}^{J+1}\omega_{j}=1$ is assumed for the mixing proportions.

The proposed approach can provide more insights into the underlying model structure based on the estimated parameters. For instance, a zero estimate of $\phi$ suggests a degeneration of the NB distribution toward Poisson. A single zero-inflation estimate can achieve degeneration toward ZINB or ZIP. The proposed model reduces to classical NB or Poisson regression when none of the potential values are selected, making it also applicable to non-inflated count data.

\subsection{Computation}
We optimize (\ref{eq:obj}) using the expectation-maximization (EM) algorithm. First, we introduce a latent indicator vector $\boldsymbol{\gamma}_{i}=\left( \gamma_{i1},\gamma_{i2}, \dots,\gamma_{iJ+1}\right)^\top$ for the $i$th observation,  where $\gamma_{ij}=1$ if the $i$th observation comes from the $j$th component and $\gamma_{ij}=0$ otherwise. Then the complete-data penalized objective function with a Lagrange multiplier term is:
\begin{align}
pl_c(\boldsymbol{\theta}) =  &\sum_{i=1}^{n}\left \{   \left(  \sum_{j=1}^{J+1}\gamma_{ij} \log\left ( \omega_{j}\right) \right) +
\gamma_{iJ+1} \log \left (   f_{\text{NB}}\left( y_i;  \exp \left(\alpha+ \boldsymbol{X}_i^\top \boldsymbol{\beta} \right) ,\phi \right) \right)  \right \} \notag \\
 -n &  \lambda_{1n} \sum_{j = 1}^p \rho_{1j} \left| \beta_j  \right| - n \lambda_{2n} \sum_{j=1}^{J} \rho_{2j} \omega_j + \delta \left( \sum_{j=1}^{J+1}\omega_j-1\right), 
\label{eq:cobj}
\end{align}
where $\delta\neq 0$ is a Lagrange multiplier. 

Based on (\ref{eq:cobj}), with fixed tuning parameters $\left ( \lambda_{1n}, \lambda_{2n} \right)$, the proposed algorithm proceeds as follows. 
\begin{enumerate} 
\item[ \textbf{Initialization}:] Set $m=0$. Set the distinct count values of response $\{y_{(1)},\cdots,y_{(J)}\}$ as the potential inflated values $\{k_1,\cdots,k_J\}$ which are sorted in an ascending order, and $J$ as the number of the distinct count values. Let $\tilde{f}_j$ be the frequency of $k_j$ for the $n$ observations. Initialize $\omega_j^{(m)}=\tilde{f}_j-\tilde{f}_1/2$ for $j=1,\cdots,J$ and $\omega_{J+1}^{(m)}=1-\sum_{j=1}^J \omega_j^{(m)}$, and $\alpha^{(m)}$, $\boldsymbol{\beta}^{(m)}$ and $\phi^{(m)}$ by fitting a traditional NB regression, where $\boldsymbol{\omega}^{(m)}$, $\alpha^{(m)}$, $\boldsymbol{\beta}^{(m)}$, and $\phi^{(m)}$  are the estimators of $\boldsymbol{\omega}$, $\alpha$ $\boldsymbol{\beta}$, and $\phi$ at step $m$, respectively. As in \cite{Zou2006}, we set $\rho_{1j} = 1/ \beta_{j}^{(0)}$ as the adaptive weights for the regression parameters. In addition, as the integers with higher frequencies are more likely to be inflated, we set 
$\rho_{2j}= 1/\omega^{(0)}_j$ to impose relatively light penalization on these points, which has satisfactory performance in our numerical studies.  

\item[ \textbf{E-step}:]  Update $m = m+1$. For $i = 1,\dots, n,$ and $j=1,\dots,J$, compute
	$$
	\hat{\gamma}_{ij}^{(m)}=E_{\boldsymbol{\theta}^{(m-1)}}\left(\gamma_{ij}\right)=\left\{\begin{matrix}
	\frac{\omega_{j}^{(m-1)}}{\omega_{j}^{(m-1)}+ \omega_{J+1}^{(m-1)} \cdot  f_{\text{NB}}\left( y_i;  \exp \left(\alpha^{(m-1)}+ \boldsymbol{X}_i^\top \boldsymbol{\beta}^{(m-1)} \right) ,\phi^{(m-1)} \right) }&\textrm{ if }y_{i}=k_{j} \\ 
	0 & \quad \text{else} 
	\end{matrix}\right., $$
	and  
	$$ \hat{\gamma}_{i,J+1}^{(m)}=E_{\boldsymbol{\theta}^{(m-1)}}\left(\gamma_{i,J+1}\right)=\left\{\begin{matrix}
	\frac{\omega_{J+1}^{(m-1)} \cdot f_{\text{NB}}\left( y_i;  \exp \left(\alpha^{(m-1)}+ \boldsymbol{X}_i^\top \boldsymbol{\beta}^{(m-1)} \right) ,\phi^{(m-1)} \right)  }{\omega_{j}^{(m-1)}+\omega_{J+1}^{(m-1)} \cdot f_{\text{NB}}\left( y_i;  \exp \left(\alpha^{(m-1)}+ \boldsymbol{X}_i^\top \boldsymbol{\beta}^{(m-1)} \right) ,\phi^{(m-1)} \right)  }&\textrm{ if } y_{i}=k_{j}\\ 
	1 & \quad \text{else}
	\end{matrix}\right.. $$

\item[ \textbf{M-step}:] 
Update $\boldsymbol{\omega}^{(m)}$, $\alpha^{(m)}$, $\boldsymbol{\beta} ^{(m)} $, and $\phi^{(m)}$ by maximizing $E_{\boldsymbol{\theta}^{(m-1)}}\left(pl_c(\boldsymbol{\theta}) \right)$ with respect to $\boldsymbol{\omega}$, $\alpha$, $\boldsymbol{\beta}$, and $\phi$, separately. 
\begin{itemize}
	\item[(M1)] For $j = 1,\dots, J+1$, update
			$$
	\omega_{j}^{(m)}=\left\{\begin{matrix}
	\frac{ n \lambda_{2n} \rho _{2j}\omega_{j}^{(m-1)} \text{I} (j\leq J)-\sum_{i=1}^{n}\hat{\gamma}_{ij}^{(m)}}{\delta} & \textrm{ if } \delta\left( n \lambda_{2n} \rho _{2j}\omega_{j}^{(m-1)} \text{I} \left( j\leq J\right) -\sum_{i=1}^{n}\hat{\gamma}_{ij}^{(m)} \right)>0 \\ 
	&\\
	0& \text{else}
	\end{matrix}\right.$$
with 
$ \delta=  n \left [ \lambda_{2n} \sum_{j=1}^{J}\rho_{2j} \omega_{j}^{(m-1)}-1 \right ] $.  
More details are presented in Section A1 of the Supplementary Material \citep{minb_supp}.  

\item[(M2)] Optimize $\alpha^{(m)}$ and $\boldsymbol{\beta}^{(m)}$ using the iteratively reweighted least squares (IRLS), as presented in Algorithm S1 in Section A2 of the Supplementary Material \citep{minb_supp}.
 
\item[(M3)] Optimize $\phi^{(m)}$ with the Newton-Raphson iterative approach.
\end{itemize}

\end{enumerate}

The E-step and M-step are conducted iteratively until convergence, where convergence is concluded if $\max \left\lbrace \left|\frac{ \alpha^{(m)}-\alpha^{(m-1)} }{ \alpha^{(m-1)} }\right|, \frac{\left\| \boldsymbol{\beta}^{(m)}-\boldsymbol{\beta}^{(m-1)}\right\| }{\left\| \boldsymbol{\beta}^{(m-1)}\right\| }, \frac{\left\| \boldsymbol{\omega}^{(m)}-\boldsymbol{\omega}^{(m-1)}\right\| }{\left\| \boldsymbol{\omega}^{(m-1)}\right\| },\left|\frac{ \phi^{(m)}-\phi^{(m-1)} }{ \phi^{(m-1)} }\right|\right\rbrace < 10^{-3}$ in our numerical study. Convergence of the algorithm is observed in all of our numerical studies. We select the two tuning parameters $\lambda_{1n} $ and $\lambda_{2n} $ through a grid search using BIC. To facilitate data analysis, we have developed an R package implementing 
the proposed approach, including sample data for application, and made it publicly available at https://github.com/rucliyang/minb as well as in the Supplementary Materiel \citep{minb_supp}.

\section{Statistical Properties}
\label{sec:asym}

We consider the scenario where both the numbers of potential inflated values $J$ and covariates $p$ are finite as the sample size increases. Denote $\phi_0, \alpha_0, \boldsymbol{\beta}_0, $ and $\boldsymbol{\omega}_0$ as the true parameters. Without loss of generality, we rewrite $\boldsymbol{\omega}_0 = \left( \boldsymbol{\omega}_{0,1}, \boldsymbol{\omega}_{0,2} \right)^{\top}$, where $\boldsymbol{\omega}_{0,1}=\left(\omega_{01}, \ldots, \omega_{0r}, \omega_{0 J+1}\right)$ consists of all the nonzero components of $\boldsymbol{\omega}_0$ and $ \boldsymbol{\omega}_{0,2}=\left(\omega_{0(r+1)}, \ldots, {\omega}_{0J}\right) $ consists of all the zero components, with $r$ being the true number of the inflated values. Similarly, denote $\boldsymbol{\beta}_0= \left(\boldsymbol{\beta}_{0,1}, \boldsymbol{\beta}_{0,2}\right)^{\top}$, where $\boldsymbol{\beta}_{0,1}$ consists of the $s$ nonzero components of $\boldsymbol{\beta}_0$ and $\boldsymbol{\beta}_{0,2}$ contains the $p-s$ zero components. As a result, the true parameters can be represented as $\boldsymbol{\theta}_0=\left( \boldsymbol{\theta}_{0,1},\boldsymbol{\theta}_{0,2} \right) = \left( \phi_0,  \alpha_0, \boldsymbol{\beta}_{0,1}, \boldsymbol{\omega}_{0,1},0,\ldots,0 \right)^\top$, where $\boldsymbol{\theta}_{0,1}$ and $\boldsymbol{\theta}_{0,2}$ contain the nonzero and zero parameters, respectively. Denote $\hat{\boldsymbol{\theta} }  = \left( \hat{ \boldsymbol{\theta}}_1, \hat{\boldsymbol{\theta}}_2 \right) $ as the maximizer of  (\ref{eq:obj}). Consider the following regularity conditions.
\begin{enumerate}
	\item[\textbf{R1}]  The density function $f\left(y_i ; \boldsymbol{X}_i, \boldsymbol{\theta}\right)$ has a common support and is identifiable with respect to $\boldsymbol{\theta}$. Moreover, it satisfies:
$$
\begin{aligned}
 \left.E\left[\frac{\partial \log f\left(y_i ; \boldsymbol{X}_i, \boldsymbol{\theta}\right)}{\partial \theta_j}\right]\right|_{\boldsymbol{\theta}=\boldsymbol{\theta}_0}&=0,  \text{and} \\ 
 E\left[\frac{\partial \log f\left(y_i ; \boldsymbol{X}_i, \boldsymbol{\theta}\right)}{\partial \theta_j} \frac{\partial \log f\left(y_i, \boldsymbol{X}_i, \boldsymbol{\theta}\right)}{\partial \theta_k}\right] & = E\left[-\frac{\partial^2 \log f\left(y_i ; \boldsymbol{X}_i, \boldsymbol{\theta}\right)}{\partial \theta_j \partial \theta_k}\right] .
\end{aligned}
$$
\item[\textbf{R2}] The Fisher information matrix for $\boldsymbol{\theta}$
$$
\mathcal{I} \left(\boldsymbol{\theta}\right)=E\left\{\left[\frac{\partial \log f\left(y_i ; \boldsymbol{X}_i, \boldsymbol{\theta}\right)}{\partial \boldsymbol{\theta}}\right]\left[\frac{\partial \log f\left(y_i ; \boldsymbol{X}_{i},\boldsymbol{\theta}\right)}{\partial \boldsymbol{\theta}}\right]^\top \right\}
$$
is finite and positive-definite at $\boldsymbol{\theta}=\boldsymbol{\theta}_0$.

	\item[\textbf{R3}] There exists an open set $\mathcal{Q}_0$ that contains the true parameter $\boldsymbol{\theta}_0$, such that for almost all  $V_i=\left(y_i, \boldsymbol{X}_i \right)$, the density $f\left(y_i ; \boldsymbol{X}_{i},\boldsymbol{\theta}\right)$  admits all third order derivatives with respect to $\boldsymbol{\theta}$ for all $\boldsymbol{\theta} \in \mathcal{Q}_0 $. Also, assume that there exist functions $B_{jkl}  \left(V_i\right)$  for all 
$j, k$, and $l$ such that 
$$ 
\left|\frac{\partial^3}{\partial \theta_j \partial \theta_k \partial \theta_l} \log f\left(y_i ; \boldsymbol{X}_i, \boldsymbol{\theta}\right)\right| \leq B_{jkl}\left(V_i \right),
$$
where there exists a constant $M$ such that $E\left[B_{jkl}  \left(V_i\right)\right]<M$.

	\item[\textbf{R4}]  $ \sqrt{n}\min \left \{ \lambda_{1n}, \lambda_{2n}  \right\} \to \infty, \liminf _{n \to \infty} \lim \inf _{\beta_j \to 0+}$ $p_{\lambda_{1n}}^{\prime}\left( \beta_j \right) / \lambda_{1n}>0 $ and \\
	$\liminf _{n \to \infty} \lim \inf _{\omega_j \to 0+}$ $p_{\lambda_{2n}}^{\prime}\left(\omega_j\right) / \lambda_{2n}>0. $  
 
\end{enumerate}

Conditions R1 to R3 are standard regularity conditions commonly assumed in published mixture model studies \citep{khalili2013regularization,Cai2014Selection,Zhong2023}. In recent mixture model studies \citep{tabrizi2020identifiability}, identifiability has been demonstrated to be still a wide-open problem. As stated in \cite{khalili2007variable}, the identifiability of the finite mixture models depends on several factors, such as component densities and the design matrix, and many finite mixture models can have identifiability under some specific conditions. We refer to the aforementioned publications for detailed discussions and sufficient conditions on identifiability. Following the existing studies, such as \cite{Cai2014Selection} and \cite{Zhong2023}, we assume that the proposed MINB model is identifiable. Condition R4 further restricts tuning parameters $\lambda_{1n}$ and $\lambda_{2n}$ with respect to the sample size $n$ and puts constraints on the penalty terms such that the nonzero coefficients can be distinguished from zero.

\begin{theorem}
\label{th1} Under conditions R1-R3, there exists a local maximizer $\hat{\boldsymbol{\theta}}$ of  $pl_n(\boldsymbol{\theta})$ such that:
$$
    \left \| \hat{\boldsymbol{\theta}}-\boldsymbol{\theta}_0 \right \|=O_{p} \left( n^{-\frac{1}{2}} + q_n \right), 
$$
where $ q_n = \max \left\{  \max \limits_{1 \leq j \leq s}\left\{ \left| p_{\lambda_{1n}}^{\prime}\left( \beta_{0j} \right) \right| ;  \beta_{0j} \neq 0\right\},  \max \limits_{1 \leq j \leq r }\left\{ p_{\lambda_{2n}}^{\prime}\left(\omega_{0j}\right); \omega_{0j} \neq 0 \right\} 
 \right\}.  $ \end{theorem}

\begin{theorem}
\label{th2}
	Under conditions R1-R4, if $q_n = O\left (  n ^{- \frac{1}{2} } \right)$,  $\hat{\boldsymbol{\theta}}  = \left( \hat{\boldsymbol{\theta}}_1, \hat{\boldsymbol{\theta}} _2 \right)$ achieves selection consistency in that $P\left( \hat{\boldsymbol{\theta}}_{2} = \boldsymbol{0} \right) \to 1 $.
\end{theorem}

Proofs of Theorems 1 and 2 are provided in Section A3 of the Supplementary Material \citep{minb_supp}. Theorem \ref{th1} shows that $\hat{\boldsymbol{\theta}}$ can achieve the desired root-$n$ convergence rate when $q_n = O\left(n ^{- \frac{1}{2} } \right)$. Theorem \ref{th2} establishes selection consistency for both the regression coefficients and the mixture proportions of inflated values. 


\section{Analysis of Online Chauffeured Service Demand Research Data}
\label{sec:analysis}

\subsection{Inflated Chauffeured Service Data}

The largest shared mobility platform in China has connected 13 million drivers and 377 million users, forming a transportation platform with 81\% of the domestic market share and a total revenue of over RMB 140 billion in 2020. In 2015, the platform launched its chauffeured (designated driving) service, committed to offering car owners convenient, professional, and reliable driving services. The OCSD survey data was collected with the purpose of better understanding service demand through data analysis and providing decision support for personalized quality services, high standards of operational management, and targeted market development. The data was collected from December 2018 to March 2019 from participants of China’s first-tier cities (i.e., Beijing, Shanghai, Guangzhou, and Shenzhen) and 97 small and medium-sized cities (e.g., Chengdu, Shenyang, etc.). After data cleaning, 3,996 complete and valid responses were received. The target variable is the number of chauffeured service usage during the past six months. The data also includes demographic, drinking behavior, and vehicle status information. A brief description of the covariates is presented in Table \ref{variable}.

\begin{table}[!ht]
      \renewcommand\arraystretch{0.95}
	\caption{ Descriptions of the covariates in OCSD data set}
	\label{variable}
\begin{center}
		\resizebox{0.9\textwidth}{!}{
		\begin{tabular}{ll}
			\hline
			\multicolumn{1}{l}{Variables}          &	\multicolumn{1}{l}{Description} \\ \hline
			Gender (Ge)                             &  1 if male, 0 if female       \\
			Age                                    &    \\
			\quad Reference                              &   18 $\sim$ 30                            \\
			\quad Prime (PA)                              &   30 $\sim$  40                                 \\
			\multicolumn{1}{c}{\quad Middle or Old (MOA)} &     $>$ 40                                       \\
			Housewife (Hw)                          &     1 if housewife, 0 otherwise                                    \\
			Work Time (WT)                           &     1 if shorter than 60 hours a week, 0 otherwise                                  \\
			Drinking                               &             \\
			\quad Reference                              &   never drinking                           \\
			\quad Seldom (SeD)                           &     lower than 2 times drinking a month      \\
			\quad Slight (SlD)                           &     2 times a week                                   \\
			\quad Moderate (MD)                           &     3 or 4 times a week         \\
			\quad Serious (SrD)                           &    5 or 6 times a week      \\
			\quad Severe (SvD)                            &     1 or more times a day    \\
			Car Price (CP)                         &     1 if higher than 500,000, 0 otherwise                                         \\
			Car Status (CS)                        &       1 if new, 0 if second-hand       \\
			Car touching (Ct)                      &       1 if not allow strangers to touch the car, 0 otherwise      \\
			Punishment                             &             \\
			\quad Reference                          &  think that drunk driving punishment is almost  non-existent     \\
			\quad Slight (SlP)                            &    think that drunk driving punishment is slight         \\
			\quad Severe (SeP)                            &     think that  drunk driving punishment is severe   \\ \hline
		\end{tabular}
	}
\end{center}
\end{table}

\begin{figure}[htbp]
	\begin{center}
		\includegraphics[width=\textwidth]{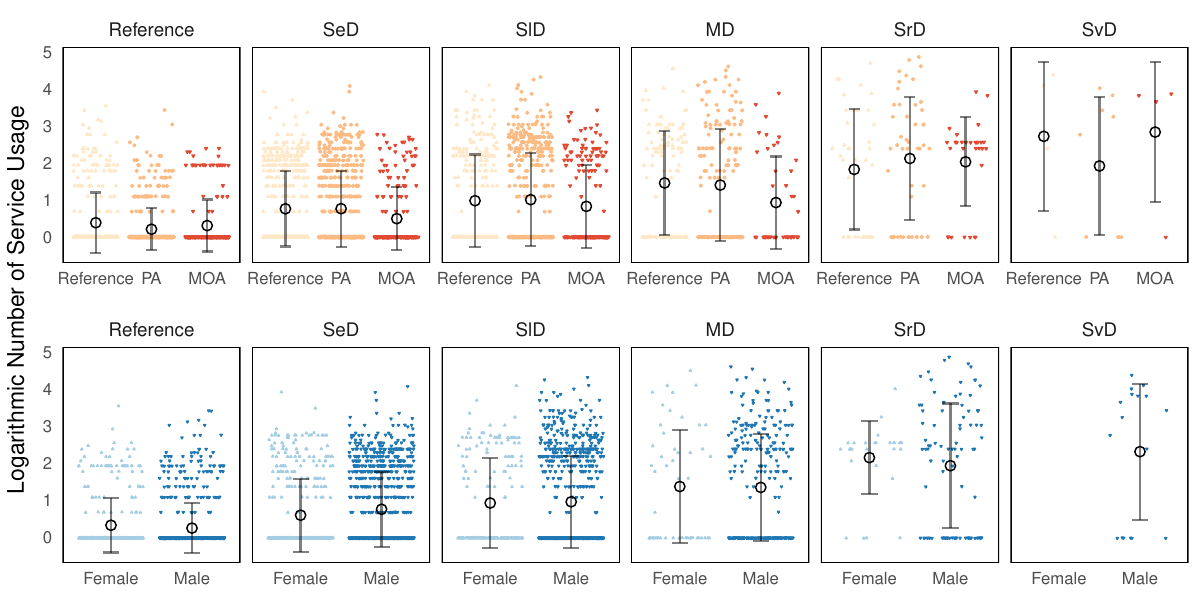}
	\end{center}
	\caption{Jitter plots of the logarithmic number of service usage by drinking frequency, age, and gender groups. The vertical error bars provide the corresponding mean $\pm$ standard error exhibited at the end of the vertical lines.}
	\label{fig:jitter}
\end{figure}

To get a deeper insight into the OCSD data, we first conduct an exploratory analysis. In Figure \ref{fig:jitter}, we provide the jitter plots with error bars (means plus and minus standard errors), which display the logarithmic number of response distributions by drinking frequency, age, and gender groups. It is observed that the average numbers of service usage in the groups with higher drinking frequencies are obviously larger than those with low frequencies. Differences and similarities in service usage patterns across different genders and age groups are also observed. However, it is difficult to identify the influencing mechanisms from the marginal descriptive analysis. On the other hand, as shown in Figure \ref{freq-original}, there is a large fraction of zero counts at six-month service usage, probably because of the heterogeneous and underreporting effects. The distribution also displays noticeable inflation at several integers, such as 10 and 20. 

To accommodate the aforementioned unique characteristics of the OCSD data, we conduct analysis with the proposed MINB approach to simultaneously identify inflated values and important covariates. In addition to the proposed approach, six alternatives, including MIP, Heaping ZIP, NB-LASSO, Poisson-LASSO, NB, and Poisson, are applied for comparison. Among them, MIP \citep{2019li} is a two-stage inflated value selection approach, and the Heaping ZIP model \citep{wang2008modeling} extends the ZIP model to the special data structure with rounded-off counts at multiples of 5, 10, and 20. Poisson and NB regression models are two popular and fully developed methods for modeling count data, and NB-LASSO and Poisson-LASSO take a further step and conduct variable selection with the LASSO penalty. 

\subsection{Results}

\begin{figure}[htbp]
	\begin{center}
		\includegraphics[width=\textwidth]{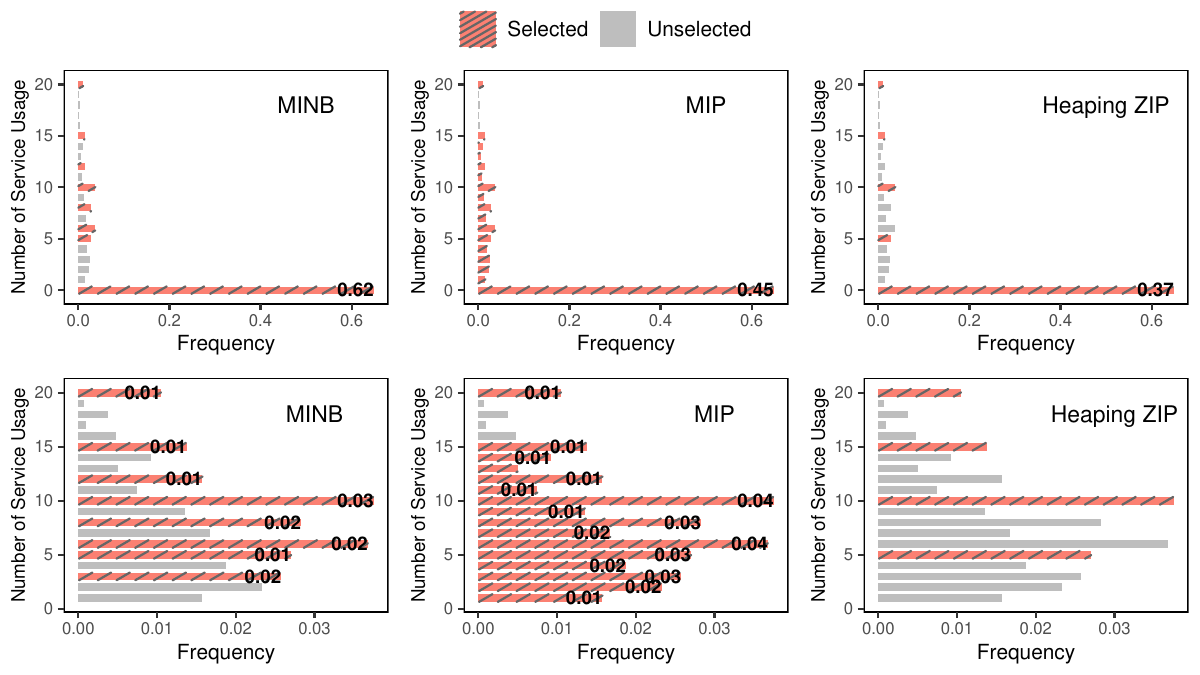}
	\end{center}
	\caption{Estimations of mixing proportions for the selected inflated values. It is noted that for Heaping ZIP, although multiple inflated values, including 0, 5, 10, 15, and 20, are considered, only the mixing proportion for zero inflation is available due to the assumed inflation mechanism. The second row provides the corresponding zoomed-in version with the x-axis from 0 to 0.04.}\label{ddomega}
\end{figure}

In Figure \ref{ddomega}, for the three approaches (proposed, MIP, and Heaping ZIP) that can accommodate inflated values, we first present the frequency distributions of the response, along with the estimated mixing proportions for the identified inflated values. The MINB approach identifies 0, 3, 5, 6, 8, 10, 12, 15, and 20 as inflated values. It leads to estimated inflated values with more practical implications compared to MIP and Heaping ZIP. In particular, with the proposed approach, the high estimated mixing proportion of 0 indicates that there exists a non-demanding group that does not utilize driving service, posing significance for the potential consumption propensity analysis. A possibility of underreporting in self-reported data is present as well. Inflated points are identified at 5, 10, 15, and 20, which is potentially due to the fact that people tend to round the count to multiples of 5 or 10. Another three inflated values are identified at 3, 6, and 12, which are not observed as inflated values in Figure \ref{freq-original}. This may result from the fact that the study investigates the frequency of service usage in the past six months, and people tend to approximate using once every two months or once/twice a month when they cannot recall the exact numbers. However, MIP selects noticeably more inflated values, probably to account for the overdispersion, which cannot be explained well by Poisson regression. In addition, Heaping ZIP only takes into account the accumulation at certain rounded-off integers, and this is less practical compared to the proposed approach.

\begin{figure}[!ht]
\begin{center}
	\includegraphics[width=\textwidth]{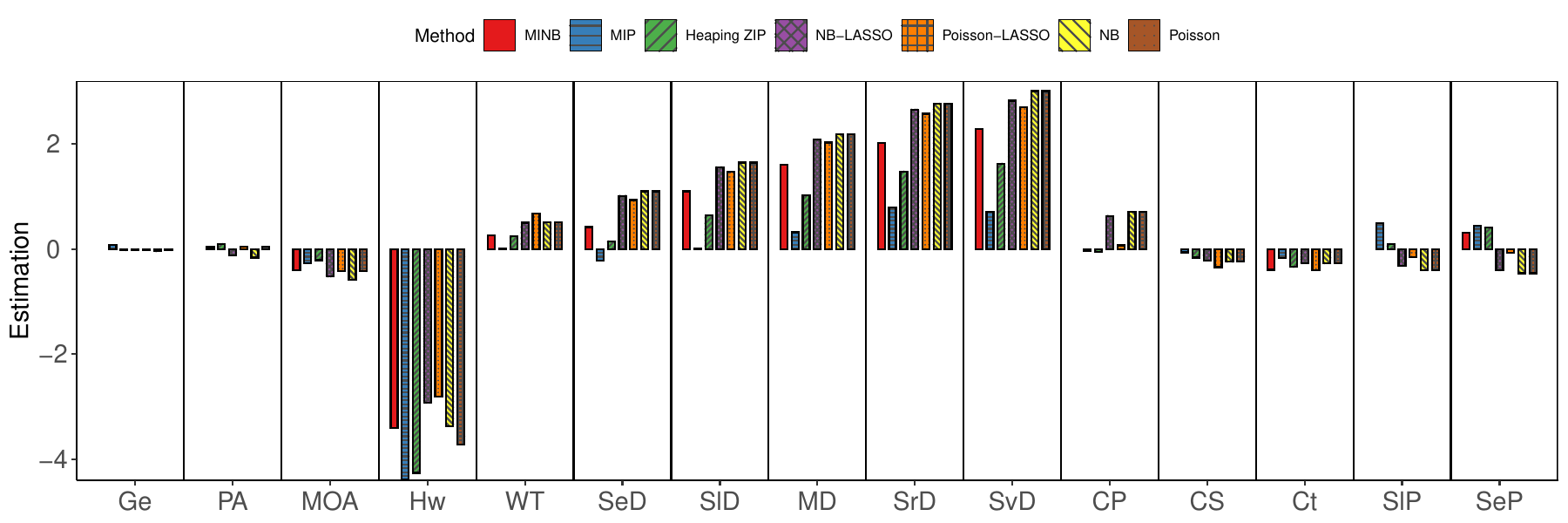}		
\end{center}
	\caption{ Estimations of regression parameters.}\label{ddbeta}
\end{figure}

The proposed approach identifies 10 important covariates, while all 15 covariates are recognized as important by NB-LASSO and Poisson-LASSO. Figure \ref{ddbeta} exhibits the estimations of regression parameters by the proposed approach and six alternatives. Four approaches, that rely on classical NB and Poisson regression and cannot account for inflation, provide estimates irreconcilable with common sense. For example, the estimated coefficients of SIP and SeP with these four approaches are all negative, suggesting that stricter government punishment toward drunk driving becomes a barrier to service demand, which is not sensible. Contrarily, the MINB model suggests that those more aware of the punishment for drunk driving are more likely to order chauffeured service. A similar conclusion has been reached in the literature. For example, \cite{hansen2015punishment} offered evidence concerning the positive effects of punishments and sanctions in reducing drunk-driving behaviors, making it possible for customers to choose chauffeured service as a safe modality. Moreover, MINB reveals a markedly significant age gap in service demand, where those younger than 40 are more likely to be the target customers of chauffeured services. This is understandable because the tech-based chauffeured service mode and its reliance on electronic equipment may increase users’ perception of the operating process complexity and thus hamper adoption for older people. There is supportive evidence in the literature \citep{sharma2012gender}, which has demonstrated that demographic age has moderating effects on online service usage and evaluation and that an easy-to-use application can improve performance expectancy and adoption intention for the advanced age groups. In addition, the proposed approach suggests that people who do not permit strangers to use their cars are reluctant to employ a designated driver. On the other hand, the differences in chauffeured service demand by the grade and newness of cars are not supported by the proposed MINB model.

Based on these results, platforms should simplify operation and decrease the complexity of application. Ease of use can augment users’ experiences, especially for the elder groups, improving their adoption. Future planning strategies may also include adequate measures to increase users’ awareness of the consequences of drunk driving. As for potential users, which often constitute high drinking frequency groups, platforms can carry out a variety of ways to attract users, such as recommendations in friend circles, advertisements, reward schemes for new users, etc.

\begin{figure}[htbp]
	\begin{center}
		\includegraphics[width=0.9\textwidth]{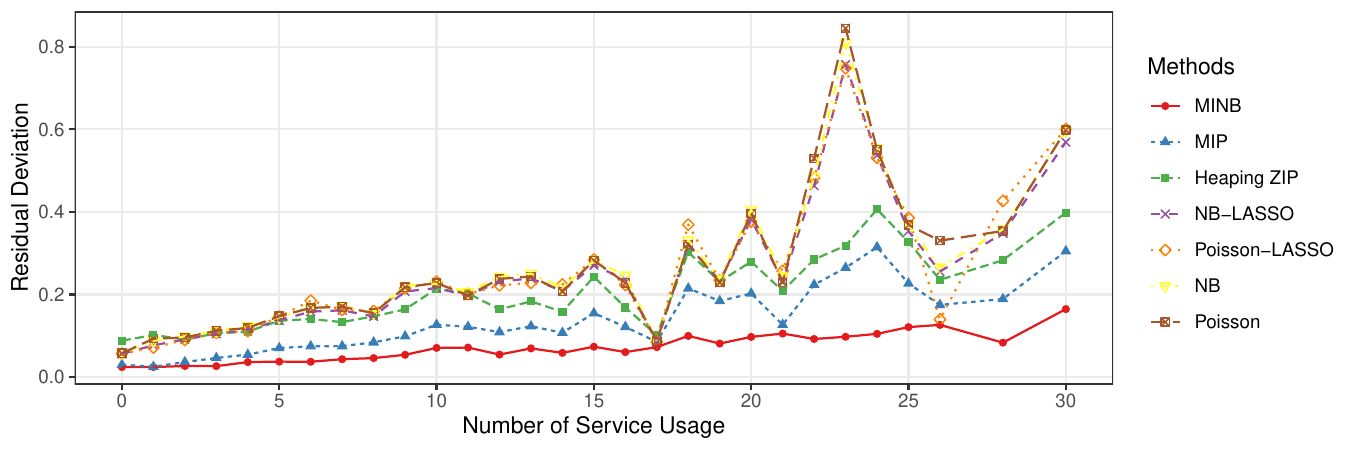}
	\end{center}
	\caption{Standard deviations of the corrected Pearson residuals. The response measure ranges from 1 to 30.}\label{ResiSD}
\end{figure}

In practical data analysis, it is difficult to objectively evaluate the inflated value and covariate identification performance, as knowledge of the ground truth is not available. To provide an indirect assessment, we first calculate the corrected Pearson residual for each approach, which is defined as $
\text{P}_{i}=\frac{O_{i}- E_{i} }{\sqrt{\hat{\psi} E_{i}  }}$ with $O_i=y_i$ being the observed value and $E_i$ being the expected value (e.g., $E_i= \sum_{j=1}^{J}\hat{\omega}_{j}k_j + \hat{\omega}_{J+1} \exp \left(\hat{\alpha}+ \boldsymbol{X}_i^\top \hat{\boldsymbol{\beta}} \right)$ for the proposed approach). 
Here, an adjustment parameter $\hat{\psi}=\frac{1}{n-d_{f}} \sum_{i=1}^{n} \frac{ \left( O_{i}- E_i\right) ^2 }{E_i} $ with $d_{f}$ being the model's degree of freedom is introduced to eliminate the influence of overdispersion and unequal variance. For each $y_{(i)} \in \{y_{(1)},\cdots,y_{(J)}\}$ (the set that includes the distinct count values of response), we compute the standard deviation of the corrected Pearson residuals of the observations with $y_{(i)}$ in Figure \ref{ResiSD}, to test for the stability of the competing approaches. As shown in Figure \ref{ResiSD}, the deviation is observably large for the NB-LASSO, Poisson-LASSO, NB, and Poisson approaches, which cannot accommodate inflation. Such a phenomenon is more obvious at the count values of 10, 15, and 20, which are selected as inflated points by the proposed approach. This suggests that a lack of appropriate accommodation of the inflated values may bias estimation and increase variation. Conversely, the inflation effect in estimation error is substantially decreased by MIP and MINB, not only at some particular integers but also with an overall decrease in the deviations, which supports the validity of the strategy for accommodating multiple inflated values.

\begin{figure}[htbp]
\begin{center}
\includegraphics[width=\textwidth]{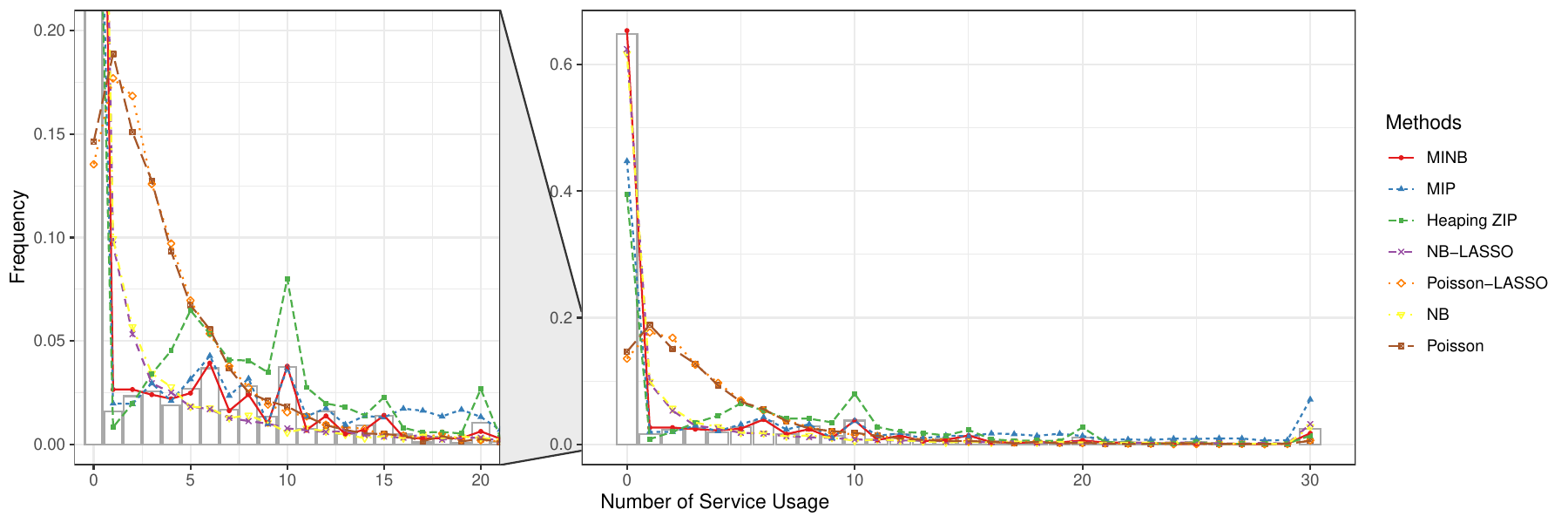}
\end{center}
\caption{Frequency distributions of the fitted results. To improve presentation, the outcome is truncated with a threshold value of 30. Left small plot: zoomed-in version with the x-axis from 1 to 20 and the y-axis from 0.00 to 0.20.}\label{ddfit}
\end{figure}

In Figure \ref{ddfit}, we further present the frequency distributions of the fitted results as well as the frequency distribution of the observed count response values. It is observed that the proposed approach provides superior performance with a properly fitted density curve at certain points. For example, the frequency of the fitted zero value is much closer to the observed data. Additionally, the fitted results at values such as 10 and 20 accurately match the true observed frequency bars. Vuong’s test is conducted to provide a more objective evaluation. Specifically, when comparing two models, a positive value of Vuong's statistic favors the former model over the latter one, with a larger value indicating more significance. The values of Vuong’s statistic for the proposed approach are 16.221 $\left(p <0.001 \right)$, 9.305 $\left(p <0.001 \right)$, 15.055 $\left(p <0.001 \right) $, 23.530 $\left(p <0.001 \right) $, 15.029 $\left(p <0.001 \right) $, and 22.811 $\left(p <0.001 \right)$ when compared to MIP, Heaping ZIP, NB-LASSO, Poisson-LASSO, NB, and Poisson, respectively. The favorable estimation and fitting performance provides support for the validity of the proposed approach.

\section{Simulation}
\label{sec:simu}
To evaluate the performance of the proposed approach, simulation studies are designed. Specifically, Simulation 1 conducts general numerical studies that cover a wide spectrum with various model settings. To further assess the validity of the proposed approach, a real-data-based simulation that mimics the OCSD data is examined in Simulation 2.

To evaluate identification performance, we adopt the true and false positive rates for covariates $\left( \text{TPR:C and} \, \text{FPR:C} \right)$ and inflated values $\left(\text{TPR:I and } \text{FPR:I} \right)$. Estimation performance is evaluated using the root sum of squared errors (RSSE:C and RSSE:I) defined as $\left\|\left(\hat{\alpha},\hat{\boldsymbol{\beta}}^{\top}\right)-\left(\alpha_0,\boldsymbol{\beta}_0^{\top}\right)\right\|_2$ and $\left\|\hat{\boldsymbol{\omega}}-\boldsymbol{\omega}_0\right\|_2$ for covariates (and intercept) and inflated variables, where $(\hat{\alpha},\hat{\boldsymbol{\beta}}, \hat{\boldsymbol{\omega}})$ and $\left(\alpha_0,\boldsymbol{\beta}_0, \boldsymbol{\omega}_0\right)$ are the estimated and true values of $\left(\alpha, \boldsymbol{\beta}, \boldsymbol{\omega} \right)$, respectively. For the dispersion parameter $\phi$, absolute error (AE:D) $|\hat{\phi}-\phi_0|$ is computed, where $\hat{\phi}$ and $\phi_0$ are the estimated and true values of $\phi$. To provide a more intuitive comparison, in addition to the six alternatives adopted above, we consider an Oracle MINB model, assuming that the true non-zero regression coefficients and inflated values are known.

\subsection{Simulation 1}

Set $n=500, 1000, 1500, $ and $2000$, and $p = 15$. The covariates are independently generated from the Normal distribution $N(0,0.5^2)$. To evaluate the performance of the proposed approach, the true models are considered to be MINB, ZINB, MIP, ZIP, and Poisson, respectively, and the intercept term $\alpha=-2$. 15 scenarios are considered, comprehensively covering different levels of sparsity and signal strength of regression parameters, diverse numbers of inflated points, and different levels of inflation.
 
Specifically, scenario 1 is the benchmark with NB parameters $\boldsymbol{\beta} = (3,1, -0.5, $ $2, -2, 2, 1,$ $ -1, 0.5, -0.5, 0,0,0,0,0)^\top$ and $ \phi=1$. The inflated values $\boldsymbol{K}$ are $ \left \{ 0,1,3, 5, 10 \right \} $ with the inflated proportions $\boldsymbol{\omega} = \left( 0.3,0.05,0.05,0.01,0.01\right)^\top$. For the dispersion parameter $\phi$, a smaller value of 0.5 and a larger value of 2 are considered in scenarios 2 and 3, respectively. For multiple inflated values, a higher level of inflation $\boldsymbol{\omega}=\left( 0.5,0.1,0.1,0.02,0.02 \right)^\top$ and a lower number of inflated values $\boldsymbol{K} = \left \{0, 1, 3 \right \}  $ with proportions $ (0.3, 0.1, 0.1)^ \top $are considered in scenarios 4 and 5, respectively. For regression parameters, $\boldsymbol{\beta} $ is set with a stronger signal $(3,1, 0.5, 2, -2, $ $2, -2, 1, 1,0, 0, 0, 0, 0 )^\top$ in scenario 6. Moreover, we consider no sparsity in scenario 7 with $\boldsymbol{\beta} = (3,1, -0.5, 2, -2, 2, 1, -1, 0.5,$ $ -0.5, 1, 1, 1, 1, 1 )^\top$. Scenario 8 considers a single inflated value of ${0}$, with an inflated proportion of $0.45$. Scenarios 9 to 14 are set with the same parameters $\boldsymbol{\beta},\boldsymbol{K},$ and $\boldsymbol{\omega}$ as those in scenarios 3 to 8, respectively, except that the true model is changed from NB to Poisson. In scenario 15, the true model is the classical Poisson with the same $\boldsymbol{\beta}$ considered in scenario 1 and no inflation.

For each scenario, 100 replicates are simulated, and the means and standard deviations (SDs) of the evaluation measures under scenario 1 are provided in Table \ref {tab:scenario1}. The remaining results for scenarios 2 to 15 are presented in Tables S1-S14 in Section A4 of the Supplementary Material \citep{minb_supp}. The proposed approach is observed to have superiority or at least competitive performance across all scenarios. For example, under scenario 1 in Table \ref{tab:scenario1}, the proposed approach satisfactorily identifies more true positives with fewer false positives for both regression parameters and inflation values with ($\text{TPR:C}, \text{FPR:C}, \text{TPR:I}, \text{FPR:I})= (0.68, 0.09, 0.92, 0.00) $ when $n = 500$, compared to $(\text{TPR:C}, \text{FPR:C})$ = (0.10, 0.00) for NB-LASSO and (0.54, 0.14) for Poisson-LASSO, and  $(\text{TPR:I}, \text{FPR:I})$ = (0.98, 0.12) for MIP.  It also performs better in estimation, with for example $ \left (\text{RSSE:C}, \text{RSSE:I}, \text{AE:D} \right ) =(1.90, 0.17, 1.34) $ when $n=500$, compared to (3.88, 0.20, -) for MIP, (3.45, 0.55, -) for Heaping ZIP, (5.53, -,8.40) for NB-LASSO, (4.19, - , -) for Poisson-LASSO, (3.43, - ,3.31) for NB, and (5.20, - , -) for Poisson. The superiority of the proposed approach becomes more evident as the sample size increases. Under the setting with $n = 4000$, the proposed approach can achieve performance comparable to the Oracle approach.

\begin{table}[htbp]
\caption{Simulation results under scenario 1 with different sample sizes. In each cell, mean(SD) based on 100 replicates.}
\setlength{\tabcolsep}{2pt}
\renewcommand\arraystretch{0.95}
\label{tab:scenario1}
\begin{tabular}{lccccccc}
\hline
Approach        & \multicolumn{1}{c}{RSSE:C}  & \multicolumn{1}{c}{TPR:C} & \multicolumn{1}{c}{FPR:C} & \multicolumn{1}{c}{ RSSE:I} & \multicolumn{1}{c}{TPR:I} & \multicolumn{1}{c}{FPR:I}  & \multicolumn{1}{c}{AE:D} \\ \hline
              & \multicolumn{7}{c}{$n =$ 500}                                                                 \\
proposed      & 1.90(0.55)  & 0.68(0.13) & 0.09(0.13) & 0.17(0.10) & 0.92(0.13) & 0.00(0.02) & 1.34(1.27)   \\
MIP           & 3.88(4.47)  &-\,-        &-\,-        & 0.20(0.10) & 0.98(0.07) & 0.12(0.08) &-\,-          \\
Heaping ZIP   & 3.45(0.60)  &-\,-        &-\,-        & 0.55(0.02) &-\,-        &-\,-        &-\,-          \\
NB-LASSO      & 5.53(0.23)  & 0.10(0.03) & 0.00(0.00) &-\,-        &-\,-        &-\,-        & 8.40(2.51) \\
Poisson-LASSO & 4.19(0.79)  & 0.54(0.22) & 0.14(0.20) &-\,-        &-\,-        &-\,-        &-\,-          \\
NB            & 3.43(0.39)  &-\,-        &-\,-        &-\,-        &-\,-        &-\,-        &  3.31(0.89) \\
Poisson       & 5.20(7.35)  &-\,-        &-\,-        &-\,-        &-\,-        &-\,-        &-\,-          \\
Oracle        & 1.09(0.34)  &-\,-        &-\,-  & 0.08(0.05) &-\,-        &-\,-  & 0.43(0.26)   \\ \hline
              & \multicolumn{7}{c}{$n =$ 1000}                                                                 \\
proposed      & 1.09(0.33)  & 0.84(0.11) & 0.10(0.14) & 0.07(0.05) & 1.00(0.02) & 0.00(0.01) & 0.41(0.35)   \\
MIP           & 2.16(1.74)  &-\,-        &-\,-        & 0.21(0.07) & 1.00(0.00) & 0.13(0.07) &-\,-          \\
Heaping ZIP   & 3.27(0.38)  &-\,-        &-\,-        & 0.55(0.01) &-\,-        &-\,-        &-\,-          \\
NB-LASSO      & 5.58(0.15)  & 0.09(0.00) & 0.00(0.00) &-\,-        &-\,-        &-\,-        & 9.02(2.03) \\
Poisson-LASSO & 3.73(0.72)  & 0.65(0.17) & 0.16(0.19) &-\,-        &-\,-        &-\,-        &-\,-          \\
NB            & 3.35(0.35)  &-\,-        &-\,-        &-\,-        &-\,-        &-\,-        &  3.49(0.69) \\
Poisson       & 5.38(6.34)  &-\,-        &-\,-        &-\,-        &-\,-        &-\,-        &-\,-          \\
Oracle        & 0.72(0.20)  &-\,-        &-\,-  & 0.06(0.03) &-\,-        &-\,-  & 0.26(0.18)   \\ \hline
              & \multicolumn{7}{c}{$n =$ 2000}                                                                \\
proposed      & 0.68(0.23)  & 0.92(0.09) & 0.08(0.12) & 0.04(0.03) & 1.00(0.00) & 0.00(0.00) & 0.22(0.19)   \\
MIP           & 1.51(0.49)  &-\,-        &-\,-        & 0.22(0.06) & 1.00(0.00) & 0.10(0.04) &-\,-          \\
Heaping ZIP   & 3.18(0.25)  &-\,-        &-\,-        & 0.55(0.01) &-\,-        &-\,-        &-\,-          \\
NB-LASSO      & 5.62(0.16)  & 0.09(0.02) & 0.00(0.00) &-\,-        &-\,-        &-\,-        & 9.39(1.92) \\
Poisson-LASSO & 3.51(0.69)  & 0.70(0.15) & 0.14(0.19) &-\,-        &-\,-        &-\,-        &-\,-          \\
NB            & 3.41(0.25)  &-\,-        &-\,-        &-\,-        &-\,-        &-\,-        & 3.71(0.47)  \\
Poisson       & 5.78(13.09) &-\,-        &-\,-        &-\,-        &-\,-        &-\,-        &-\,-          \\
Oracle        & 0.48(0.14)  &-\,-        &-\,-  & 0.04(0.03) &-\,-        &-\,-  & 0.18(0.13)   \\ \hline
              & \multicolumn{7}{c}{$n =$ 4000}                                                                \\
proposed      & 0.41(0.15)  & 0.98(0.05) & 0.06(0.10) & 0.03(0.02) & 1.00(0.00) & 0.00(0.00) & 0.14(0.11)   \\
MIP           & 1.30(0.35)  &-\,-        &-\,-        & 0.22(0.04) & 1.00(0.00) & 0.08(0.03) &-\,-          \\
Heaping ZIP   & 3.19(0.18)  &-\,-        &-\,-        & 0.55(0.01) &-\,-        &-\,-        &-\,-          \\
NB-LASSO      & 5.63(0.11)  & 0.09(0.00) & 0.00(0.00) &-\,-        &-\,-        &-\,-        & 9.38(1.05) \\
Poisson-LASSO & 3.27(0.57)  & 0.73(0.12) & 0.10(0.16) &-\,-        &-\,-        &-\,-        &-\,-          \\
NB            & 3.39(0.18)  &-\,-        &-\,-        &-\,-        &-\,-        &-\,-        & 3.82(0.30)  \\
Poisson       & 5.47(2.62)  &-\,-        &-\,-        &-\,-        &-\,-        &-\,-        &-\,-          \\
Oracle        & 0.33(0.09)  &-\,-        &-\,-  & 0.03(0.02) &-\,-        &-\,-  & 0.12(0.10)   \\ \hline
\end{tabular}
\end{table}

In general, the performance of the proposed approach appears to be stable when changing the dispersion degree (Tables S1 and S2), number and mixing proportions of true inflated values (Tables S3 and S4), signal level (Table S5), and sparsity level (Table S6). Similar patterns are observed, where the proposed approach outperforms the alternatives in terms of all evaluation measures. Even under scenarios 8 to 15, where the data generation model is misspecified, the proposed approach still provides performance comparable to the approach along with the true model.

\begin{figure}[htbp]
	\centering		\includegraphics[width=0.81\textwidth]{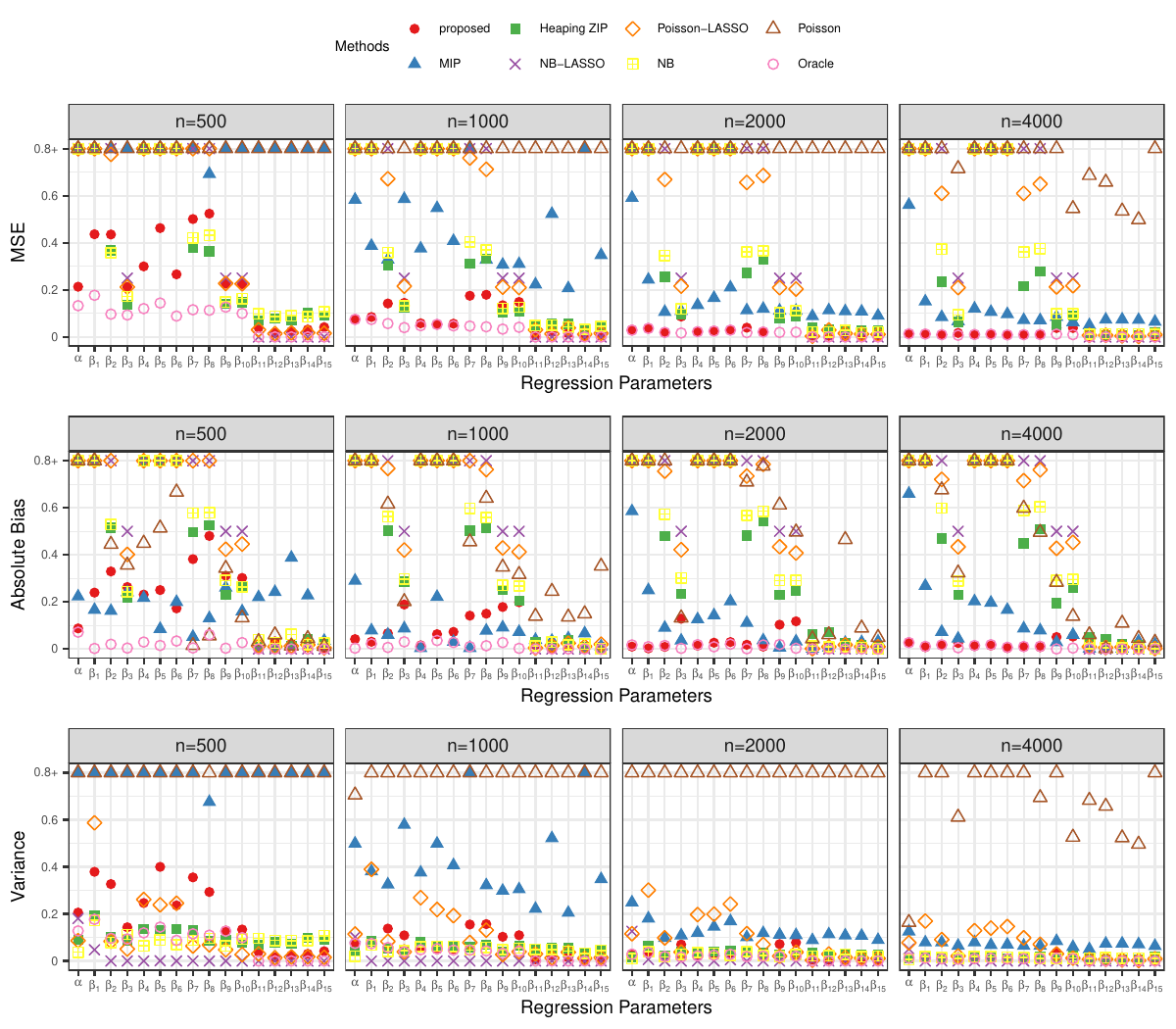}
	\caption{MSE, absolute bias, and variance values of regression parameters over 100 replicates, under scenario 1 with different sample sizes.} \label{fig:sim_mse_beta}
\end{figure}

\begin{figure}[htbp]
	\centering
		\includegraphics[width=0.81\textwidth]{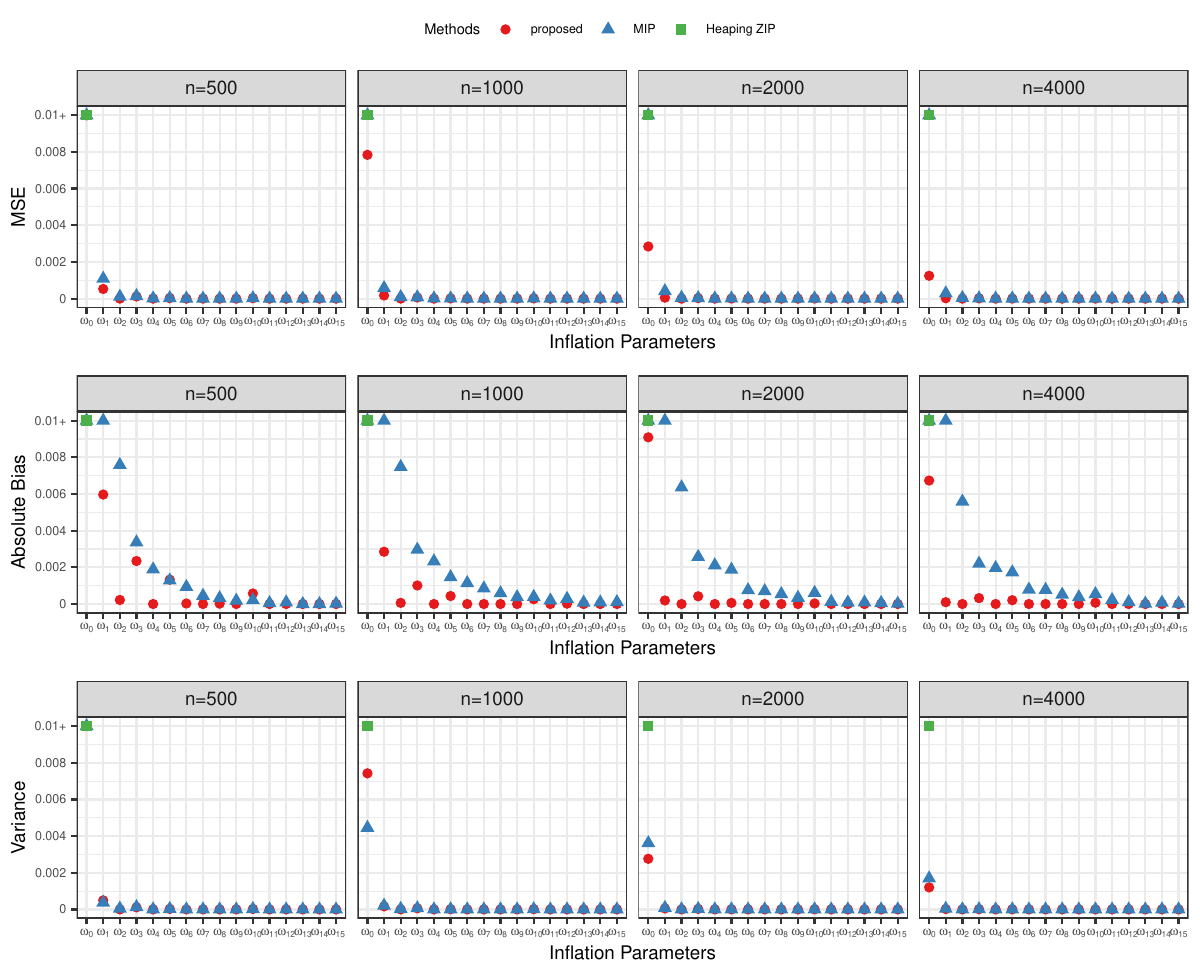}
	\caption{MSE, absolute bias, and variance values of inflation parameters over 100 replicates, under scenario 1 with different sample sizes.} \label{fig:sim_mse_omega}
\end{figure}

To provide a more lucid demonstration, we examine the detailed estimation results under scenario 1. The summarized absolute bias, variance, as well as MSE values for the regression coefficients and inflated proportions are provided in Figures \ref{fig:sim_mse_beta} and \ref{fig:sim_mse_omega}, respectively. In Figure \ref{fig:sim_mse_beta}, it is observed that when the sample size is relatively small, the proposed approach has a slightly larger MSE than the competitors for a few parameters with weak signals, such as $\beta_3, \beta_7, $ and $\beta_8$. This result is expected because the proposed model is more intricate when inflated values are accommodated, and the estimated performance is not so stable with variance somehow increasing. However, with an increasing sample size, the MSEs of the proposed approach have an obvious decrease and show sizable advantages over the competing approaches. As for inflated values, in Figure \ref{fig:sim_mse_omega}, the proposed approach is observed to provide more satisfactory estimations of $\boldsymbol{\omega}$ compared to MIP and Heaping ZIP (only at the zero point), with a noticeably smaller bias, especially at integers with relatively large inflated proportions such as 0 and 3. In addition, the variance is again observed to decrease markedly with the increased sample size. 

\subsection{Simulation 2}

We further conduct a real data-based simulation. The simulated settings are based on the specific covariate structure of the OCSD data. In particular, for each replicate, 80\% of the subjects are randomly sampled from the OCSD data without replacement, and we use the observed covariate measurements. The response $y_i$ is generated from the MINB model (\ref{eq:model}), where the true parameters $\alpha$ and $\boldsymbol{\beta}$ are set as 2 and $(1.34,0,0, -0.40, -3.40, 0.26, 0.42, 1.10,1.60,2.02,2.29,0,0,-0.39,0,0.31)^\top$ to mimic the real regression scheme. The dispersion parameter $\phi$ is accordingly set as 1. In addition, nine inflated values $\{0,3,5,6,8,10,12,15,20\}$ are considered, and the corresponding mixing proportions are set according to the estimates in Section \ref{sec:analysis} as $\left( 0.62,0.02,0.01,0.02,0.02,0.03,0.01,0.01,0.01\right)^\top$. 

Table \ref{tab:realdata_sim} reports the summarized simulation results based on 100 replicates. The proposed approach is observed to show better identification accuracy for regression parameters, where it selects the true signals with a higher TPR of 0.80 and lower FPR of 0.07, compared to (0.67, 0.09) for NB-LASSO and (0.73, 0.24) for Poisson-LASSO. For the inflated values, the proposed approach also maintains superiority in that it can identify the majority of the true inflated points (TPR = 1.00) with almost no false positives (FPR = 0.00). It again achieves better estimation performance with (RSSE:C, RSSE:I, AE:D)=(3.40, 0.01, 0.12) compared to (8.50, 0.06, -) for MIP, (3.78,0.62, -) for Heaping ZIP, (4.30, - ,5.94) for NB-LASSO, (4.11, - , -) for Poisson-LASSO, (4.14, - ,5.83) for NB, and (11.54, - , -) for Poisson. Heaping ZIP achieves the second-best estimation performance, perhaps because the true inflated values are mostly set as multiples of 5, which is consistent with the model assumption. The real data-based numerical studies additionally reveal that the proposed approach can help improve model performance on both identification and estimation.

\begin{table}[htbp]
\caption{Simulation results under the real data-based scenario. In each cell, mean (SD) based on 100 replicates.}
\setlength{\tabcolsep}{2pt}
\renewcommand\arraystretch{1}
\label{tab:realdata_sim}
\begin{tabular}{lccccccc}
\hline
Approach       & \multicolumn{1}{c}{RSSE:C}  & \multicolumn{1}{c}{TPR:C} & \multicolumn{1}{c}{ FPR:C} & \multicolumn{1}{c}{RSSE:I} & \multicolumn{1}{c}{TPR:I} & \multicolumn{1}{c}{FPR:I}  & \multicolumn{1}{c}{AE:D} \\ \hline
proposed      & 3.40(1.58)  & 0.80(0.11) & 0.07(0.15) & 0.01(0.01) & 1.00(0.02) & 0.00(0.01) & 0.12(0.09) \\
MIP           & 8.50(8.48)  &-\,-         &-\,-         & 0.06(0.00) & 1.00(0.00) & 0.27(0.04) &-\,-         \\
Heaping ZIP   & 3.78(1.19)  &-\,-         &-\,-         & 0.62(0.00) &-\,-         &-\,-         &-\,-         \\
NB-LASSO      & 4.30(0.24)  & 0.67(0.09) & 0.09(0.15) &-\,-         &-\,-         &-\,-         & 5.94(0.27) \\
Poisson-LASSO & 4.11(0.26)  & 0.73(0.08) & 0.24(0.20) &-\,-         &-\,-         &-\,-         &-\,-         \\
NB            & 4.14(5.39)  &-\,-         &-\,-         &-\,-         &-\,-         &-\,-         & 5.83(0.26) \\
Poisson       & 11.54(7.45) &-\,-         &-\,-         &-\,-         &-\,-         &-\,-         &-\,-         \\
Oracle        & 2.81(5.16)  &-\,-  &-\,- & 0.01(0.01) &-\,-  &-\,-  & 0.09(0.07) \\
\hline
\end{tabular}
\end{table}

\section{Discussion}
\label{sec:disc} 
The inflation issue in investigation data poses an analytical concern in market research. In addition to zero inflation, there is also the phenomenon of inflation at multiple values. In this study, we have developed a novel approach to identify important influencing factors on count response, with the impact of multiple inflated values effectively accommodated as well. The proposed model is more general with an extra dispersion parameter; it includes MIP, ZINB, ZIP, NB, and Poisson as special cases. The statistical properties have been rigorously established where the estimator has been proven to have satisfactory estimation and selection consistency properties. The effectiveness of the proposed approach has been further verified by extensive simulation studies.

The proposed approach has been applied to influencing factor analysis with the OCSD data. Superior performance in terms of both estimation and identification has been demonstrated by comprehensive quantitative and visualized diagnosis. It has been shown that with adequate accommodation of multiple inflated values, the proposed approach presents a significant advantage over the alternative approaches. By identifying the important influencing factors on designated driving service demand, a series of valuable conclusions have been drawn. The proper interpretation and implications can further serve as useful guidance and decision support for the platform. The model is potentially applicable to diverse fields, such as social psychology \citep{socialpsy2020}, personality and individual differences \citep{personality2017}, and public health \citep{health2017}.

This study can be potentially extended in multiple aspects. The generalized multiple-inflated NB model has been adopted. Although it includes multiple models as special cases, it still might be sensitive to deviations from the assumed distribution. Recently, new approaches based on generalized estimating equations (GEE) have been proposed to provide distribution-free alternatives for count data regression with variable selection \citep{chen2016variable,chen2022distribution}. Further investigation on adopting estimating equations for more robust analysis will be pursued in future work. We have considered the scenario where there are multiple unknown inflated values and focused on estimating covariates-response associations and identifying inflated values, which is much more challenging than existing studies that have only focused on investigating covariates-response associations with prior known inflated values. We have assumed covariate-independent mixing probabilities to facilitate a less complicated formulation and theoretical proof. Such an assumption has also been adopted in \cite{wang2008modeling}, \cite{crawford2015sex}, and \cite{yee2022generally}. It can be of interest to extend the proposed approach to accommodate covariate-dependent mixing probabilities. This extension is expected to be nontrivial and may demand a separate development. In this study, we have identified important covariates based on the estimated regression coefficients via the penalization technique as opposed to statistical inference. Inference under the penalization framework has been demonstrated to be nontrivial. It will be more challenging with the proposed analysis as we have identification on both inflated values and covariates. We have assumed that the subjects are independent. In applied studies, complex mechanisms often exist, such as network structures, that may influence service demand. Investigating multiple-inflated count data analysis under complex relationships is another topic for future research.

\begin{acks}[Acknowledgments]
The authors would like to thank the anonymous referees, an Associate
Editor, and the Editor for their constructive comments which led to a significant improvement of this article.

\end{acks}
%
\begin{funding}
The first author was supported by National Natural Science Foundation of China (72271237) and MOE Project of Key Research Institute of Humanities and Social Sciences (22JJD910001). The third author was supported by National Natural Science Foundation of China 12071273.
\end{funding}

\begin{supplement}

\stitle{Supplementary A to “Identification of Influencing Factors on Self-reported Count Data with Multiple Potential Inflated Values”} 
\sdescription{We provide the analytical solution for $\boldsymbol{\omega}^{(m)}$ in Section A1 and the IRLS algorithm for $\alpha^{(m)}$ and $\boldsymbol{\beta}^{(m)}$ in  Section A2. The proofs of Theorems 1 and 2 are presented in Section A3. Section A4 provides the additional simulation results (Tables S1-S14). } 
\end{supplement}
\begin{supplement}

\stitle{Supplementary B to “Identification of Influencing Factors on Self-reported Count Data with Multiple Potential Inflated Values”}
\sdescription{R code for implementing the proposed approach as well as sample data for application are provided as a supplement and can also be found at https://github.com/rucliyang/minb.}
\end{supplement}


\bibliographystyle{imsart-nameyear} 
\bibliography{ref}       

\end{document}


\begin{frontmatter}
\title{ Supplementary A to “Identification of Influencing Factors on Self-reported Count Data with Multiple Potential Inflated Values”}
\runtitle{ MINB}

%
%

\end{frontmatter}

\section{Analytical Solution for the $\boldsymbol{\omega}^{(m)}$ Optimization}

The first-order condition of the weights is expressed as: 
\begin{equation}
\label{eq:s5}
\frac{\partial pl_c\left( \boldsymbol{\theta};\boldsymbol{\gamma}^{(m)}\right) }{\partial \omega_j} =\sum_{i=1}^{n}\frac{\hat{\gamma}_{ij}^{(m)}}{\omega_{j}^{(m-1)}}
- n \lambda_{2n} \rho_{2j} \text{I} \left( j\leq J\right) +\delta =0. 
\end{equation}

By multiplying $\omega_{j}^{(m-1)}$ on both sides of Equation (\ref{eq:s5}) , one has:
\begin{equation}
\label{eq:s6}
\sum_{i=1}^{n}\hat{\gamma }_{ij}^{(m)}-  n \lambda_2\rho_{2j} \omega_{j}^{(m-1)}+\delta \omega_{j}^{(m-1)}=0. 
\end{equation}

Summing over (\ref{eq:s6}) and by $\sum_{j=1}^{J+1}\omega _{j}^{(m-1)}=1$, we have:

$$
 \delta=  n \left [ \lambda_{2n} \sum_{j=1}^{J}\rho_{2j} \omega_{j}^{(m-1)}-1 \right ].
 $$
 
Plugging $\delta$ back to (\ref{eq:s6}) gives:
	$$
	\omega_{j}^{(m)}=\left\{\begin{matrix}
	\frac{ n \lambda_{2n} \rho _{2j}\omega_{j}^{(m-1)} \text{I} (j\leq J)-\sum_{i=1}^{n}\hat{\gamma}_{ij}^{(m)}}{\delta} & \quad\text{ if }\left( n \lambda_{2n} \rho _{2j}\omega_{j}^{(m-1)} \text{I} \left( j\leq J\right) -\sum_{i=1}^{n}\hat{\gamma}_{ij}^{(m)} \right) \cdot\delta>0 \\ 
	&\\
	0& \text{else}.
	\end{matrix}\right.$$

\clearpage
\section{IRLS Algorithm for the $\alpha^{(m)}$ and $\boldsymbol{\beta}^{(m)}$ Optimization}

\quad

\begin{algorithm}[H]
 \renewcommand{\algorithmicrequire}{\textbf{Input:}}
 \renewcommand{\algorithmicensure}{\textbf{Output:}}
 \caption{ Optimize $\alpha^{(m)}$ and $\boldsymbol{\beta}^{(m)}$ at step $m$ by IRLS}
 \begin{algorithmic}[1]
  \REQUIRE $X_{\cdot j}, \hat{\gamma}_{\cdot,J+1 }^{(m)}, \alpha^{(m-1)}, \boldsymbol{\beta}^{(m-1)},$ and $ \phi^{(m-1)} $
  \ENSURE  $ \alpha^{(q)} $ as $ \alpha^{(m)} $, and
  $\boldsymbol{\beta}^{(q)}$ as $ \boldsymbol{\beta}^{(m)}$
  \STATE Set $q = 0$. Initialize $\alpha^{(q)} = \alpha^{(m-1)},$ and $\boldsymbol{\beta}^{(q)} = \boldsymbol{\beta}^{(m-1)}$.
  \WHILE { $ \max \left\{ \left| \alpha^{(q)}-
\alpha^{(q-1)} \right|, \left\| \boldsymbol{\beta}^{(q)} - \boldsymbol{\beta}^{(q-1)}\right\| \right\}  \geq 10^{-3}$ } 
\STATE Update $q = q+1 $
  \STATE Compute  $\omega_{i0}^{(q)} =\frac{\left(y_{i}-\mu_{i}^{(q-1)}\right)}{\left(1 + \mu_{i}^{(q-1)}\phi^{(m-1)}\right)}  +  \frac{-2y_i \phi^{(m-1)} \mu_{i}^{(q-1)}+\phi^{(m-1)}\left(\mu^{(q-1)}_{i}\right)^2-y_i}
	 { \left(1 + \mu_{i}^{(q-1)}\phi^{(m-1)}\right) ^2}  $ and $\tau_{i0}^{(q)} = \alpha^{(q-1)} -  \frac{\left(y_{i}-\mu_{i}^{(q-1)}\right)}{\left(1 + \mu_{i}^{(q-1)}\phi^{(m-1)}\right) \omega_{i0}^{(q)} },$ with $\mu_{i}^{(q-1)}= \text{exp}\left( \alpha^{(q-1)} + \sum_{j=1}^p \beta_{j}^{(q-1)}X_{ij} \right).$ 
  \STATE Update $ \alpha^{(q)} = \frac{\sum_{i} \hat{\gamma}_{i,J+1}^{(m)} \omega_{i0}^{(q)}\tau_{i0}^{(q)} 
	 }{\sum_{i} \hat{\gamma}_{i,J+1}^{(m)} \omega_{i0}^{(q)}}.$
  \FOR{$j \gets 1$ to $p$}   
  \STATE Update $\mu_{i}^{(q)}= \text{exp}\left( \alpha^{(q)} + \sum_{k=1}^{j-1} \beta_{k}^{(q)}X_{ik} +  \sum_{k=j}^p \beta_{k}^{(q-1)}X_{ik}\right).$ 
   \STATE Compute $\omega_{ij}^{(q)} =\frac{\left(y_{i}-\mu_{i}^{(q)}\right)}{\left(1 + \mu_{i}^{(q)}\phi^{(m-1)}\right)} X_{ij}^2 +  \frac{-2y_i \phi^{(m-1)} \mu_{i}^{(q)}+\phi^{(m-1)}\left(\mu^{(q)}_{i}\right)^2-y_i}
	 { \left(1 + \mu_{i}^{(q)}\phi^{(m-1)}\right) ^2}  X_{ij}^2$, and $\tau_{ij}^{(q)} = \beta_{j}^{(q-1)} -  \frac{\left(y_{i}-\mu_{i}^{(q)}\right)}{\left(1 + \mu_{i}^{(q)}\phi^{(m-1)}\right) \omega_{ij}^{(q)} } X_{ij}.$ 
	
	\STATE Update $\beta_{j}^{(q)} = \text{argmax}\left[ \frac{1}{2}\sum_{i=1}^{n}  \hat{\gamma}_{i,J+1}^{(m)} \omega_{ij}^{(q)}\left(\tau_{ij}^{(q)} - \beta_{j}\right)^2 - n  \lambda_{1n} \rho_{1j} \left|\beta_{j} \right|\right] $
	with the  solution:
	\begin{align*}
	    	\beta_{j}^{(q)} = & \operatorname{sign}\left(\tilde{\beta}_{j}\right) \\
	&\left[ \operatorname{sign}\left(\tilde{\beta}_{j}\right) \left( \frac{\sum_{i} \hat{\gamma}_{i,J+1}^{(m)} \omega_{ij}^{(q)}\tau_{ij}^{(q)} 
	 + n \lambda_{1n} \rho_{1j} \operatorname{sign}\left(\tilde{\beta}_{j} \right)}{\sum_{i} \hat{\gamma}_{i,J+1}^{(m)} \omega_{ij}^{(q)}}\right) \right]_{+},
	\end{align*}
	where $\tilde{\beta}_{j} = \frac{\sum_{i} \hat{\gamma}_{i,J+1}^{(m)} \omega_{ij}^{(q)}\tau_{ij}^{(q)} 
	 }{\sum_{i} \hat{\gamma}_{i,J+1}^{(m)} \omega_{ij}^{(q)}}$ and $f_{+}=max(f,0)$. 
  \ENDFOR
  \ENDWHILE
\end{algorithmic}
\end{algorithm}

\clearpage
\section{Proofs of Theorems}
\begin{proof}[Proof of Theorem 1]
	Consider $b_n= n^{-\frac{1}{2}} + q_n  $. It is sufficient to show that for any given $\epsilon$, there exists a constant $C_\epsilon$, such that:
	\begin{equation}
	\label{eq:s1}
	 \lim_{n \to \infty}P\left \{ \sup_{\left \| \boldsymbol{u} \right \|=C_\epsilon} pl_n\left( \boldsymbol{\theta}_0 +b_n\boldsymbol{u}\right) <pl_n\left( \boldsymbol{\theta}_0 \right) \right \}\geq 1-\epsilon.  
	\end{equation}
	Therefore, with probability of at least $1-\epsilon$, there is a local maximum in $\left\lbrace \boldsymbol{\theta}_0 + b_n\boldsymbol{u};\left\| \boldsymbol{u}\right\| \leq C_\epsilon\right\rbrace$ satisfying $\left \| \hat{\boldsymbol{\theta}}-\boldsymbol{\theta}_0 \right \|=O_{p}\left( b_n\right)$.
	
Denote $d_{\boldsymbol{\theta}}$ and $t$ as the dimension of the parameter space and the number of true parameters, respectively. Namely, $d_{\boldsymbol{\theta}} = p + J + 3 $ and $t = r + s + 3$.	Consider $D_n(\boldsymbol{u})=pl_{n}\left( \boldsymbol{\theta}_0 +b_n\boldsymbol{u}\right) -pl_n\left( \boldsymbol{\theta}_0 \right) $. By using $p_{\lambda_{ \cdot n}}(\cdot)>0$, we have:
	\begin{align}
	\label{eq:s2}
	D_n(\boldsymbol{u})&=\left [ l_n \left(\boldsymbol{\theta}_0 +b_n\boldsymbol{u} \right) -  n \sum_{j = 1}^p p_{\lambda_{1n}} \left(\beta_j  \right) - n \sum_{j = 1}^J p_{\lambda_{2n}} \left( \omega_j \right)  \right ] - 
	 \left [  l_n(\boldsymbol{\theta}_0 )- n \sum_{j = 1}^s p_{\lambda_{1n}} \left( \beta_{0j}  \right) - n\sum_{j=1}^{r}p_{\lambda_{2n}}(\omega_{0j})\right ]  \notag \\
	 & = \left [ l_n \left(\boldsymbol{\theta}_0 +b_n\boldsymbol{u} \right) -  n \sum_{j = 1}^s p_{\lambda_{1n}} \left( \beta_j \right) - n \sum_{j = s+1}^p p_{\lambda_{1n}} \left( \beta_j  \right)  - n \sum_{j = 1}^r p_{\lambda_{2n}} \left( \omega_j \right)  - n \sum_{j = r+1}^J p_{\lambda_{2n}} \left( \omega_j \right)  \right ]  \notag \\
	 & -    \left [  l_n(\boldsymbol{\theta}_0 )- n \sum_{j = 1}^s p_{\lambda_{1n}} \left( \beta_{0j}   \right) - n\sum_{j=1}^{r}p_{\lambda_{2n}}(\omega_{0j})\right ] \notag \\
	&\leq \left[l_n(\boldsymbol{\theta}_0 +b_n\boldsymbol{u})- l_n(\boldsymbol{\theta}_0 )\right] -n \sum_{j=1}^{s}\left[ p_{\lambda_{1n}}( \beta_{j}  )-p_{\lambda_{1n}}( \beta_{0j}  ) \right]  -n \sum_{j=1}^{r}\left[ p_{\lambda_{2n}}(\omega_{j})-p_{\lambda_{2n}}(\omega_{0j}) \right] \notag \\
	&  \leq  \left[l_n(\boldsymbol{\theta}_0 +b_n\boldsymbol{u})- l_n(\boldsymbol{\theta}_0 ) \right] + n  \left | \sum_{j=1}^{s}   \left 
[ p_{\lambda_{1n}} \left(  \beta_{j}   \right )-p_{\lambda_{1n}} \left( \beta_{0j} \right) \right] \right |   + n  \left |   \sum_{j=1}^{r}   \left [ p_{\lambda_{1n}} \left(\omega_j  \right )-p_{\lambda_{1n}} \left(  \omega_{0j} \right) \right] \right |     \notag \\
	&=D_{1n}(\boldsymbol{u})+D_{2n}(\boldsymbol{u}) + D_{3n}(\boldsymbol{u}).
		\end{align}
	
	For $D_{1n}(\boldsymbol{u})$ , by applying Taylor expansion, we obtain:
	\begin{align*}
	D_{1n}(\boldsymbol{u})&=b_nl_n'\left( \boldsymbol{\theta}_0 \right) ^\top \boldsymbol{u}+\frac{b_{n}^{2}}{2}\boldsymbol{u}^\top l_n''\left( \boldsymbol{\theta}_0\right) \boldsymbol{u}+\frac{b_n^3}{6}[\boldsymbol{u}^\top\partial l''_n(\boldsymbol{\theta}^* )/\partial \boldsymbol{\theta}\boldsymbol{u}]\boldsymbol{u}\\
	&=D_{1,1n}+D_{1,2n}+D_{1,3n},
	\end{align*}
	where $l_n'(\cdot)$ and $l_n''(\cdot)$ are the first and second derivatives of $l_n(\cdot)$, and $\boldsymbol{\theta}^* $ is between $\boldsymbol{\theta}_0$ and $\boldsymbol{\theta}_0+b_n\boldsymbol{u}$. 
	With condition R1, we have: 
\begin{align*}
\left.\frac{1}{\sqrt{n}} \frac{\partial l_n ' \left(\boldsymbol{\theta} \right)}{\partial \boldsymbol{\theta}}\right|_{\boldsymbol{\theta}_0} & =\sqrt{n}\left(\left.\frac{1}{n} \sum_{i=1}^n \frac{\partial \log f(y_i ; \boldsymbol{X}_i,\boldsymbol{\theta}) }{\partial \boldsymbol{\theta}}\right|_{\boldsymbol{\theta}_0}\right) \\
& =\sqrt{n}\left(\left.\frac{1}{n} \sum_{i=1}^n \frac{\partial \log f(y_i ; \boldsymbol{X}_i,\boldsymbol{\theta}) }{\partial \boldsymbol{\theta}}\right|_{\boldsymbol{\theta}_0}-\left.E\left[\frac{\partial \log f \left( y; \boldsymbol{X}, \boldsymbol{\theta} \right)}{\partial \boldsymbol{\theta} }\right]\right|_{\boldsymbol{\theta}_0}\right)  \\ 
& \rightarrow N(0, \tilde{\Sigma}).
\end{align*}
Thus  $ l_n'(\boldsymbol{\theta}_0)= O_p\left( \sqrt{n}\right)$. Then, 
$$	
	\left | D_{1,1n} \right|  =O_p\left( \sqrt{n}b_n\right) \left \| \boldsymbol{u} \right \|  \leq O_p\left( nb_n^2\right) \left\|\boldsymbol{u}\right\|. 
$$
	For $D_{1,2n}$, 
	\begin{align*}
	D_{1,2n}&=\frac{b_{n}^{2}}{2}\boldsymbol{u}^\top E\left( l_n''(\boldsymbol{\theta}_0)\right) \boldsymbol{u}+\frac{b_{n}^{2}}{2}\boldsymbol{u}^\top \left[ l_n''(\boldsymbol{\theta}_0)-E\left( l_n''(\boldsymbol{\theta}_0)\right) \right] \boldsymbol{u}. 
		\end{align*}
For any $\varepsilon$, by Chebyshev's inequality,
\begin{align*}
& P\left(\left\|\frac{1}{n} \left\{  l_n''(\boldsymbol{\theta}_0)-E\left( l_n''(\boldsymbol{\theta}_0)\right) \right\}  \right\| \geq \varepsilon \right) \\
& \quad \leq \frac{1}{n^2 \varepsilon^2} E \sum_{j, k}^{d_{\boldsymbol{\theta}}}\left\{\frac{\partial ^2 l_n\left(\boldsymbol{\theta}_{ 0}\right)}{\partial \theta_{j} \theta_{k}}-E \frac{\partial^2 l_n\left(\boldsymbol{\theta}_{0}\right)}{\partial \theta_{j} \theta_{k}}\right\}^2 \\
& =\frac{d_{\boldsymbol{\theta}}^2}{n \varepsilon^2}=o_p(1). 
\end{align*}
Thus, $ \left\|\frac{1}{n} \left\{  l_n''(\boldsymbol{\theta}_0)-E\left( l_n''(\boldsymbol{\theta}_0)\right) \right\}  \right\| = o_p(1) $. Then, 
$$
D_{1,2n} =-\frac{nb_{n}^{2}}{2}\boldsymbol{u}^\top\left[ \mathcal{I} (\boldsymbol{\theta}_0)\right] \boldsymbol{u}+\frac{nb_{n}^{2}}{2}\left\| \boldsymbol{u}\right\| ^2 \times o_p(1). 
$$
For $D_{1,3n}(\boldsymbol{u})$, by condition  {R3} and Cauchy-Schwartz inequality, we have: 
	\begin{align*}
	\left| D_{1,3n}\right| &=\frac{b_{n}^{3}}{6}\left|\sum_{j,k,l}^{d_{\boldsymbol{\theta}}} \frac{\partial^3l_n(\boldsymbol{\theta}^*)}{\partial \theta_j\partial \theta_k \partial \theta_l}u_ju_ku_l \right|=\frac{b_{n}^{3}}{6}\left|\sum_{j,k,l}^{d_{\boldsymbol{\theta}}} \sum_{i=1}^{n}\frac{\partial^3\log f(y_i ; \boldsymbol{X}_i,\boldsymbol{\theta}^*)}{\partial \theta_j\partial\theta_k \partial\theta_l}u_ju_k u_l \right|\\
	& \leq \frac{b_{n}^{3}}{6} \sum_{i=1}^{n} \left\lbrace \sum_{j,k,l}^{d_{\boldsymbol{\theta}}}B^2_{jkl}(\boldsymbol{v_i})\right\rbrace  ^{1/2}\left\|\boldsymbol{u} \right\|^3 =nO_p\left( d_{\boldsymbol{\theta}}^{3/2}\right)  \times b_n^3 \times \left\|\boldsymbol{u} \right\|^3 \\
	&= O_p\left( d_{\boldsymbol{\theta}}^{3/2} b_n \right) \times nb_n^2 \times \left\|\boldsymbol{u} \right\|^2.
	\end{align*}
	Since $d_{\boldsymbol{\theta}} \ll n    $, we have: 
	$$
	D_{1,3n}=o_p(nb_n^2)\left\| \boldsymbol{u}\right\|^2.  
	$$
For $D_{2n}(\boldsymbol{u})$, by Taylor’s expansion and the condition $p''_{\lambda_{1n}}(\cdot)=0$, we have:
	\begin{align*}
	D_{2n}(\boldsymbol{u})& = n   \left |    \sum_{j=1}^{s} \left[  p_{\lambda_{1n}} \left(  \beta_{j}   \right )-p_{\lambda_{1n}} \left( \beta_{0j} \right) \right] \right | \\
	& = n \left |  \sum_{j=1}^{s}  p'_{\lambda_{1n}}( \beta_{0j}  ) \left( \beta_{j} - \beta_{0j}  \right)   \right|   \\
	&\leq n \left |  \sum_{j=1}^{s} \max \limits_{j }\left(  p'_{\lambda_{1n}}( \beta_{0j}  )\right)  \left( \beta_{j} - \beta_{0j}  \right)   \right|.   
	\end{align*}
By Cauchy-Schwartz inequality, we have:
	\begin{align*}
	D_{2n}(\boldsymbol{u})& \leq  n \sqrt{s} \, \left( \max \limits_{j} \left|  p'_{\lambda_{1n}}( \beta_{0j} ) \right | \right) b_n  \left\| \boldsymbol{u}\right\| \\
	& \leq n \sqrt{s} \,q_n   b_n  \left\| \boldsymbol{u}\right\|  \\ 
	&= O_p( n b^2_n ) \left\| \boldsymbol{u}\right\|. 
	\end{align*}
	Similarly, we obtain that $D_{3n}(\boldsymbol{u}) =  O_p( n b^2_n ) \left\| \boldsymbol{u}\right\| $.
	
	By Condition  {R2}, the Fisher information matrix $ \mathcal{I} (\boldsymbol{\theta}_0)$ is positive-definite. Thus, the quadratic function 
	$$ 
	-\frac{nb_{n}^{2}}{2}\boldsymbol{u}^\top  \mathcal{I} (\boldsymbol{\theta}_0) \boldsymbol{u}
	$$
	is the sole leading term on the right-hand side of (A.2). Thus, for any given $\epsilon >0 $, by choosing a sufficiently large $C_\epsilon$, (\ref{eq:s1}) holds. This completes the proof.
\end{proof}

The result of Lemma \ref{le1} is used to prove Theorem 2. 
\begin{lemma}
\label{le1}
	Under the Conditions of Theorem 2, for any given $\boldsymbol{\theta}_1$ in the neighborhood $\left\|\boldsymbol{\theta}_1-\boldsymbol{\theta}_{0,1} \right\|=O_p\left( n ^ {-\frac{1}{2}}\right)  $ and any constant $C$, as $ n \to \infty $, with probability tending to 1,
$$
pl_n\left(\boldsymbol{\theta}_{1}, \boldsymbol{0}\right) = \max _{\left\| \boldsymbol{\theta}_{2}\right\| \leq C n ^{ - 1 / 2}}  pl_n \left(\boldsymbol{\theta}_{1}, \boldsymbol{\theta}_{2}\right).
$$
\end{lemma}
\begin{proof}[proof of Lemma \ref{le1}] 
	  Let $\varepsilon_n = C n ^{ - 1 / 2}$, it is sufficient to show that with probability tending to 1 as $ n \to \infty$, for any  $\boldsymbol{\theta}_1$ satisfying $ \left \| \boldsymbol{\theta}_1 - \boldsymbol{\theta}_{0,1} \right \| = O_p\left( n ^ {-\frac{1}{2}}\right) $, we have, for $j = t+1, \cdots, d_{\boldsymbol{\theta}}$,
\begin{align}
\frac{\partial  pl_n\left(\boldsymbol{\theta}\right)}{\partial {\theta}_{j}}<0 &  \quad \text { for } 0<\theta_{j}<\varepsilon_n, \label{eq:align1} \\
\frac{\partial  pl_n\left(\boldsymbol{\theta}\right)}{\partial {\theta}_{j}}>0 & \quad \text { for }-\varepsilon_n<\theta_{ j}<0 . 
\label{eq:align2}
\end{align}

	By applying the mean value theorem, we have:
	\begin{equation} 
	\label{eq:s3}
	l_n(\boldsymbol{\theta}_{1},\boldsymbol{\theta}_2)-l_n(\boldsymbol{\theta_{1}},\boldsymbol{0})= \left[ \frac{\partial l_n(\boldsymbol{\theta}_1,\boldsymbol{\zeta})}{\partial\boldsymbol{\theta}_2} \right]^\top\boldsymbol{\theta}_2 =\sum_{j>t}^{d_{\boldsymbol{\theta}}}\frac{\partial l_n(\boldsymbol{\theta}_1,\boldsymbol{\zeta})}{\partial\theta_j}\times \theta_j, 
	\end{equation}
	where $\boldsymbol{\zeta}$ is between $\boldsymbol{0}$ and $\boldsymbol{\theta}_2$; thus $\left\|\boldsymbol{\zeta} \right\| \leq \left\|\boldsymbol{\theta}_2 \right\| = O_p\left( n ^ {-\frac{1}{2}}\right) $. Applying triangle inequality leads to the result for each $j$,
	\begin{equation}
	\label{eq:s4}
	\left| \frac{\partial l_n(\boldsymbol{\theta}_1,\boldsymbol{\zeta})}{\partial\theta_j}-\frac{\partial l_n(\boldsymbol{\theta}_{0,1},\boldsymbol{0})}{\partial\theta_j}\right| \leq \left| \frac{\partial l_n(\boldsymbol{\theta}_1,\boldsymbol{\zeta})}{\partial\theta_j}-\frac{\partial l_n(\boldsymbol{\theta}_1,\boldsymbol{0})}{\partial\theta_j}\right| +\left| \frac{\partial l_n(\boldsymbol{\theta}_1,\boldsymbol{0})}{\partial\theta_j}-\frac{\partial l_n(\boldsymbol{\theta}_{0,1},\boldsymbol{0})}{\partial\theta_j}\right|. 
	\end{equation}
	By applying Taylor's expansion,
	\begin{align*}
	\frac{\partial l_n(\boldsymbol{\theta}_1,\boldsymbol{0})}{\partial\theta_j}-\frac{\partial l_n(\boldsymbol{\theta}_{0,1},\boldsymbol{0})}{\partial\theta_j}&=\sum_{k=1}^{t }\frac{\partial^2 l_n(\boldsymbol{\theta}_{0,1},\boldsymbol{0})}{\partial\theta_k\theta_j} \times (\theta_k-\theta_{0k})\\
	&+\sum_{k=1}^{t}\sum_{l=1}^{t}\frac{\partial^3 l_n(\boldsymbol{\zeta}^*,\boldsymbol{0})}{\partial\theta_l\theta_k\theta_j} \times (\theta_k-\theta_{0k})(\theta_l-\theta_{0l})\\
	&=(A)+(B), 
	\end{align*}
	where $\boldsymbol{\zeta}^*$ is between  $\boldsymbol{\theta}_1$ and $\boldsymbol{\theta}_{0,1}$. An asymptotic order assessment of $(A)$ and $(B)$ is as follows.

 For $(A)$, we can show that 
	\begin{align*}
	(A)&=\sum_{k=1}^{t }E\left[\frac{\partial^2  l_n(\boldsymbol{\theta}_{0,1},\boldsymbol{0})} {\partial\theta_k\theta_j}\right] \times (\theta_k - \theta_{0k})+\sum_{k=1}^{t}\left \{ \frac{\partial^2 l_n\left( \boldsymbol{\theta}_{0,1},\boldsymbol{0}\right) }{\partial\theta_k\theta_j}- E \left[ \frac{\partial^2  l_n(\boldsymbol{\theta}_{0,1},\boldsymbol{0})} {\partial\theta_k\theta_j}\right] \right \} \times (\theta_k-\theta_{0k})\\
	&=(A1)+(A2).
	\end{align*}
	We have:
	$$ 
	(A1)= \sum_{k=1}^{t }E\left[\frac{\partial^2  l_n(\boldsymbol{\theta}_{0,1},\boldsymbol{0})} {\partial\theta_k\theta_j}\right] \times (\theta_k - \theta_{0k}) =  n \sum_{k=1}^{t} - \mathcal{I}_{jk}(\boldsymbol{\theta}_{0,1},\boldsymbol{0})  \times (\theta_k - \theta_{0k}), 
	$$
where $ \mathcal{I}_{jk}(\boldsymbol{\theta}_{0,1},\boldsymbol{0}) $ is the $jk$th element of information matrix $\mathcal{I}$ evaluated at $(\boldsymbol{\theta}_{0,1},\boldsymbol{0})$. According to Cauchy-Schwarz inequality,
	$$ 
	\left|(A1) \right|\leq n \left\{ \sum_{k=1}^{t} \mathcal{I}^2_{jk}(\boldsymbol{\theta}_{0,1},\boldsymbol{0})  \right\} ^{1/2} \times \left\| \boldsymbol{\theta}_1-\boldsymbol{\theta}_{0,1}\right\|, 
	$$
	which leads to:
	$$
	\left|(A1) \right|\leq n \times O_p\left( 1\right)  \times O_p(n ^{-\frac{1}{2} } )= O_p(\sqrt{n}).
	$$
	Also, by a similar argument, 
	$$
	\left| (A2)\right| \leq   \left\lbrace \sum_{k=1}^{t}\left( \frac{\partial^2 l_n(\boldsymbol{\theta}_{0,1},\boldsymbol{0})}{\partial\theta_k\theta_j}- E\left[\frac{\partial^2  l_n(\boldsymbol{\theta}_{0,1},\boldsymbol{0})} {\partial\theta_k\theta_j}\right] \right)^2 \right\rbrace ^{1/2} \times \left\| \boldsymbol{\theta}_1-\boldsymbol{\theta}_{0,1}\right\|.
	$$
	By the regularity conditions, we have:
	\begin{align*}
	&\left( \frac{\partial^2 l_n(\boldsymbol{\theta}_{0,1},\boldsymbol{0})}{\partial\theta_k\theta_j}- E\left[\frac{\partial^2  l_n(\boldsymbol{\theta}_{0,1},\boldsymbol{0})} {\partial\theta_k\theta_j}\right] \right)^2 = \left( \sum_{i=1}^{n}\left[\frac{\partial^2 \log f \left(y_i ; \boldsymbol{X}_i, \left( \boldsymbol{\theta}_{0,1},\boldsymbol{0} \right) \right)}{\partial\theta_k\theta_j} -E\left( \frac{\partial^2 \log f \left(y_i ; \boldsymbol{X}_i, \left( \boldsymbol{\theta}_{0,1},\boldsymbol{0} \right) \right)}{\partial\theta_k\theta_j}\right) \right]  \right)^2\\
	&= n\left(\frac{1}{\sqrt n}\sum_{i=1}^{n}\left[\frac{\partial^2 \log f
 \left (y_i ; \boldsymbol{X}_i,
 \left( \boldsymbol{\theta}_{0,1},\boldsymbol{0} \right) \right)}{\partial\theta_k\theta_j} -E\left( \frac{\partial^2 \log f \left(y_i ; \boldsymbol{X}_i, \left( \boldsymbol{\theta}_{0,1},\boldsymbol{0}\right) \right)}{\partial\theta_k\theta_j}\right) \right] \right) ^2=n \times \left( O_p(1)\right) ^2=O_p(n).
	\end{align*}
	Hence, $ \left|(A2)\right| \leqslant O_p\left(n \right) ^{1 / 2}\left\|\boldsymbol{\theta}_1-\boldsymbol{\theta}_{0,1}\right\| $ , suggesting that:
	$$
	(A)=(A1)+(A2)=O_p(\sqrt{n}).
	$$
    For $(B)$,
	\begin{align*}
	\left| (B)\right|&= \left|   \sum_{k=1}^{t}\sum_{l=1}^{t}\frac{\partial^3 l_n(\boldsymbol{\zeta}^*,\boldsymbol{0})}{\partial\theta_l\theta_k\theta_j} \times (\theta_k-\theta_{0k})(\theta_l-\theta_{0l})  \right| \\
	&= \left| \sum_{i=1}^{n}\sum_{k=1}^{t}\sum_{l=1}^{t}\frac{\partial^3 \log f\left(y_i ; \boldsymbol{X}_i, \left( \boldsymbol{\zeta}^*,\boldsymbol{0}\right) \right)}{\partial\theta_l\theta_k\theta_j}(\theta_k-\theta_{0k})(\theta_l-\theta_{0l})\right| \\
	&\leq \sum_{i=1}^{n} \left\lbrace\sum_{k=1}^{t}\sum_{l=1}^{t}B^2_{jkl}(\boldsymbol{v}_i) \right\rbrace^{1/2} \times \left\| \boldsymbol{\theta}_1-\boldsymbol{\theta}_{0,1}\right\|^2\\
	&=n \times O_p(n^{-\frac{1}{2}})^2 = O_p(1).
	\end{align*}
With the order assessment of $(A)$ and $(B)$, for each $j = t+1, \dots, d_{\boldsymbol{\theta}}$,
	$$
	\frac{\partial l_n(\boldsymbol{\theta}_1,\boldsymbol{0})}{\partial\theta_j}-\frac{\partial l_n(\boldsymbol{\theta}_{0,1},\boldsymbol{0})}{\partial\theta_j} =O_p(\sqrt{n}).
	$$
	Similarly,
	$$
	\frac{\partial l_n(\boldsymbol{\theta}_1,\boldsymbol{\zeta})}{\partial\theta_j}-\frac{\partial l_n(\boldsymbol{\theta}_1,\boldsymbol{0})}{\partial\theta_j} =O_p(\sqrt{n}).
	$$
	With $ \frac{\partial l_n(\boldsymbol{\theta}_{0,1},\boldsymbol{0})} { \partial \theta_j } = O_p(\sqrt n)$ and equation (\ref{eq:s4}), we have $ \frac{ \partial l_n(\boldsymbol{\theta}_1,\boldsymbol{\zeta})} { \partial \theta_j } = O_p(\sqrt{n })$. Thus, we get:
	$$
	 \frac{ \partial pl_n(\boldsymbol{\theta}_1,\boldsymbol{\zeta})} { \partial \theta_j } = O_p(\sqrt{n }) - n P'_{\lambda_{\cdot n} }\left( \left | \theta_j \right| \right) \text{sign}( \theta_j )  = n \lambda_{\cdot n}   \left [ - \frac{P'_{\lambda_{\cdot n} } \left( \left | \theta_j \right| \right) } { \lambda_{\cdot n} }  \text{sign}( \theta_j )  + O_p \left( \frac{n^{- \frac{1}{2}}}{\lambda_{\cdot n}}\right)  \right ] 	.$$
	According to condition R4, we have $ \sqrt{n}  \lambda_{\cdot n}  \to \infty, $ and $ \liminf _{n \to \infty} \lim \inf _{\theta_j \to 0+}$ $p_{\lambda_{\cdot n}}^{\prime}\left(\theta_j\right) / \lambda_{ \cdot n}>0 $. Thus,  (\ref{eq:align1}) and  (\ref{eq:align2}) follow. This completes the proof.
\end{proof}
\begin{proof}[proof of Theorem 2]
	Consider the partition $\boldsymbol{\theta}=\left( \boldsymbol{\theta}_1,\boldsymbol{\theta}_2\right)$ in the neighborhood $\left\|\boldsymbol{\theta}-\boldsymbol{\theta}_0 \right\|=O_p(n^ {-\frac{1}{2}}) $. By assuming that$\left( \hat{\boldsymbol{\theta}}_{1},\boldsymbol{0}\right)$ is the local maximizer of the penalized log-likelihood $pl_n\left( \boldsymbol{\theta},\boldsymbol{0}\right)  $, it suffices to show that as $n \to \infty$,
	\[P\left\lbrace  pl_n (\boldsymbol{\theta}_1,\boldsymbol{\theta}_2) < pl_n (\hat{\boldsymbol{\theta}}_{1},\boldsymbol{0})\right\rbrace  \to 1.\]
	First, we have:
	\begin{align*}
	pl_n(\boldsymbol{\theta}_1,\boldsymbol{\theta}_2)-pl_n\left( \hat{\boldsymbol{\theta}}_{1},\boldsymbol{0}\right) &=pl_n(\boldsymbol{\theta}_1,\boldsymbol{\theta}_2)-pl_n(\boldsymbol{\theta}_{1},\boldsymbol{0})+pl_n(\boldsymbol{\theta}_{1},\boldsymbol{0})-pl_n(\hat{\boldsymbol{\theta}}_{1},\boldsymbol{0}).
	\end{align*}
	By the definition of $\hat{\boldsymbol{\theta}}_{1}$, for any $\boldsymbol{\theta}_1$ in the neighborhood $\left\|\boldsymbol{\theta}-\boldsymbol{\theta}_0 \right\|=O_p(n^ {-\frac{1}{2}})$, we have $pl_n(\boldsymbol{\theta}_{1},\boldsymbol{0}) \leq pl_n(\hat{\boldsymbol{\theta}}_{1},\boldsymbol{0})$. Therefore, by Lemma \ref{le1},
	$$
	pl_n(\boldsymbol{\theta}_1,\boldsymbol{\theta}_2)-pl_n(\hat{\boldsymbol{\theta}}_{1},\boldsymbol{0}) \leq pl_n(\boldsymbol{\theta}_1,\boldsymbol{\theta}_2)-pl_n(\boldsymbol{\theta}_{1},\boldsymbol{0}) <0, 
	$$
	with probability tending to 1, as $n \to \infty$. This completes the proof of Theorem 2.
\end{proof}

\clearpage
\section{Additional Simulation Results}
\subsection*{}
\begin{table}[htbp]
\caption{   Simulation results under scenario 2 with different sample sizes. In each cell, mean(SD) based on 100 replicates.}
\setlength{\tabcolsep}{1.5pt}
\renewcommand\arraystretch{1}
\resizebox{\textwidth}{!}{
\begin{tabular}{lccccccc}
\hline
Approach        & \multicolumn{1}{c}{RSSE:C}  & \multicolumn{1}{c}{TPR:C} & \multicolumn{1}{c}{FPR:C} & \multicolumn{1}{c}{ RSSE:I} & \multicolumn{1}{c}{TPR:I} & \multicolumn{1}{c}{FPR:I}  & \multicolumn{1}{c}{AE:D} \\ \hline
              & \multicolumn{7}{c}{$n =$ 500}                                                                 \\
proposed      & 1.38(0.65)   & 0.89(0.11) & 0.18(0.20) & 0.08(0.08) & 0.95(0.10) & 0.00(0.01) & 0.57(0.91)    \\
MIP           & 3.17(4.54)   & -\,-         & -\,-         & 0.10(0.07) & 0.99(0.05) & 0.12(0.09) & -\,-            \\
Heaping ZIP   & 3.49(0.52)   & -\,-         & -\,-         & 0.54(0.01) & -\,-         & -\,-         & -\,-            \\
NB-LASSO      & 6.68(2.65)   & 0.10(0.06) & 0.00(0.02) & -\,-         & -\,-         & -\,-         & 10.50(2.74) \\
Poisson-LASSO & 3.95(0.78)   & 0.84(0.13) & 0.33(0.23) & -\,-         & -\,-         & -\,-         & -\,-            \\
NB            & 3.68(0.43)   & -\,-         & -\,-         & -\,-         & -\,-         & -\,-         & 3.47(0.71)  \\
\hline
\end{tabular}}
\end{table}

\begin{table}[htbp]
\caption{   Simulation results under scenario 3 with different sample sizes. In each cell, mean(SD) based on 100 replicates.}
\setlength{\tabcolsep}{1.5pt}
\renewcommand\arraystretch{1}
\resizebox{\textwidth}{!}{
\begin{tabular}{lccccccc}
\hline
Approach        & \multicolumn{1}{c}{RSSE:C}  & \multicolumn{1}{c}{TPR:C} & \multicolumn{1}{c}{FPR:C} & \multicolumn{1}{c}{ RSSE:I} & \multicolumn{1}{c}{TPR:I} & \multicolumn{1}{c}{FPR:I}  & \multicolumn{1}{c}{AE:D} \\ \hline
              & \multicolumn{7}{c}{$n =$ 500}                                                                 \\
proposed      & 2.15(0.60)  & 0.62(0.13) & 0.08(0.12) & 0.21(0.08) & 0.91(0.14) & 0.00(0.01) & 2.37(1.60)  \\
MIP           & 5.23(5.62)  & -\,-         & -\,-         & 0.31(0.08) & 0.95(0.10) & 0.13(0.09) & -\,-           \\
Heaping ZIP   & 3.87(0.44)  & -\,-         & -\,-         & 0.58(0.03) & -\,-         & -\,-         & -\,-           \\
NB-LASSO      & 5.52(0.20)  & 0.09(0.00) & 0.00(0.00) & -\,-         & -\,-         & -\,-         & 8.64(2.62) \\
Poisson-LASSO & 4.34(0.72)  & 0.51(0.21) & 0.17(0.20) & -\,-         & -\,-         & -\,-         & -\,-           \\
NB            & 3.55(0.47)  & -\,-         & -\,-         & -\,-         & -\,-         & -\,-         & 3.44(1.12)  \\
\hline

\end{tabular}}
\end{table}

\begin{table}[htbp]
\caption{   Simulation results under scenario 4 with different sample sizes. In each cell, mean(SD) based on 100 replicates.}
\setlength{\tabcolsep}{1.5pt}
\renewcommand\arraystretch{1}
\resizebox{\textwidth}{!}{
\begin{tabular}{lccccccc}
\hline
Approach        & \multicolumn{1}{c}{RSSE:C}  & \multicolumn{1}{c}{TPR:C} & \multicolumn{1}{c}{FPR:C} & \multicolumn{1}{c}{ RSSE:I} & \multicolumn{1}{c}{TPR:I} & \multicolumn{1}{c}{FPR:I}  & \multicolumn{1}{c}{AE:D} \\ \hline
              & \multicolumn{7}{c}{$n =$ 500}                                                                 \\
proposed      & 1.89(0.57)  & 0.63(0.13) & 0.05(0.10) & 0.16(0.09) & 1.00(0.00) & 0.00(0.01) & 1.16(1.20)   \\
MIP           & 3.93(5.64)  & -\,-         & -\,-         & 0.14(0.07) & 1.00(0.00) & 0.12(0.09) & -\,-           \\
Heaping ZIP   & 3.69(0.47)  & -\,-         & -\,-         & 0.76(0.01) & -\,-         & -\,-         & -\,-           \\
NB-LASSO      & 5.43(0.27)  & 0.12(0.05) & 0.00(0.02) & -\,-         & -\,-         & -\,-         & 5.23(2.12) \\
Poisson-LASSO & 4.46(0.70)  & 0.48(0.21) & 0.12(0.18) & -\,-         & -\,-         & -\,-         & -\,-           \\
NB            & 3.70(0.29)  & -\,-         & -\,-         & -\,-         & -\,-         & -\,-         &  1.76(0.52)  \\
\hline

\end{tabular}}
\end{table}

\begin{table}[htbp]
\caption{   Simulation results under scenario 5 with different sample sizes. In each cell, mean(SD) based on 100 replicates.}
\setlength{\tabcolsep}{1.5pt}
\renewcommand\arraystretch{1}
\resizebox{\textwidth}{!}{
\begin{tabular}{lccccccc}
\hline
Approach        & \multicolumn{1}{c}{RSSE:C}  & \multicolumn{1}{c}{TPR:C} & \multicolumn{1}{c}{FPR:C} & \multicolumn{1}{c}{ RSSE:I} & \multicolumn{1}{c}{TPR:I} & \multicolumn{1}{c}{FPR:I}  & \multicolumn{1}{c}{AE:D} \\ \hline
              & \multicolumn{7}{c}{$n =$ 500}                                                                \\
proposed      & 3.66(0.89)  & 0.40(0.18) & 0.06(0.12) & 0.28(0.15) & 1.00(0.03) & 0.02(0.05) & 5.70(3.22) \\
MIP           & 6.77(7.58)  & -\,-         & -\,-         & 0.08(0.04) & 1.00(0.03) & 0.05(0.10) & -\,-              \\
Heaping ZIP   & 4.83(0.43)  & -\,-         & -\,-         & 0.93(0.04) & -\,-         & -\,-         & -\,-              \\
NB-LASSO      & 5.46(0.19)  & 0.09(0.01) & 0.00(0.00) & -\,-         & -\,-         & -\,-         & 5.35(2.14)    \\
Poisson-LASSO & 5.07(0.66)  & 0.24(0.22) & 0.05(0.14) & -\,-         & -\,-         & -\,-         & -\,-              \\
NB            & 4.56(0.37)  & -\,-         & -\,-         & -\,-         & -\,-         & -\,-         & 3.43(0.78)    \\
\hline
\end{tabular}}
\end{table}

\begin{table}[htbp]
\caption{   Simulation results under scenario 6 with different sample sizes. In each cell, mean(SD) based on 100 replicates.}
\setlength{\tabcolsep}{1.5pt}
\renewcommand\arraystretch{1}
\resizebox{\textwidth}{!}{
\begin{tabular}{lccccccc}
\hline
Approach        & \multicolumn{1}{c}{RSSE:C}  & \multicolumn{1}{c}{TPR:C} & \multicolumn{1}{c}{FPR:C} & \multicolumn{1}{c}{ RSSE:I} & \multicolumn{1}{c}{TPR:I} & \multicolumn{1}{c}{FPR:I}  & \multicolumn{1}{c}{AE:D} \\ \hline
              & \multicolumn{7}{c}{$n =$ 500}                                                                 \\

proposed      & 1.78(0.63)    & 0.83(0.11) & 0.12(0.19) & 0.14(0.10) & 0.91(0.13) & 0.00(0.00) & 1.24(1.36)    \\
MIP           & 7.57(17.38)   & -\,-         & -\,-         & 0.21(0.10) & 0.98(0.07) & 0.17(0.12) & -\,-            \\
Heaping ZIP   & 3.60(0.59)    & -\,-         & -\,-         & 0.55(0.01) & -\,-         & -\,-         & -\,-            \\
NB-LASSO      & 6.48(0.31)    & 0.09(0.00) & 0.00(0.00) & -\,-         & -\,-         & -\,-         & 11.25(2.97) \\
Poisson-LASSO & 3.95(0.81)    & 0.81(0.14) & 0.36(0.24) & -\,-         & -\,-         & -\,-         & -\,-            \\
NB            & 3.90(0.39)    & -\,-         & -\,-         & -\,-         & -\,-         & -\,-         & 3.65(0.66)  \\
\hline
\end{tabular}}
\end{table}

\begin{table}[htbp]
\caption{   Simulation results under scenario 7 with different sample sizes. In each cell, mean(SD) based on 100 replicates.}
\setlength{\tabcolsep}{1.5pt}
\renewcommand\arraystretch{1}
\resizebox{\textwidth}{!}{
\begin{tabular}{lccccccc}
\hline
Approach        & \multicolumn{1}{c}{RSSE:C}  & \multicolumn{1}{c}{TPR:C} & \multicolumn{1}{c}{FPR:C} & \multicolumn{1}{c}{ RSSE:I} & \multicolumn{1}{c}{TPR:I} & \multicolumn{1}{c}{FPR:I}  & \multicolumn{1}{c}{AE:D} \\ \hline
              & \multicolumn{7}{c}{$n =$ 500}                                                                 \\

proposed      & 1.81(0.58)   & 0.85(0.11) & -\,-         & 0.12(0.10) & 0.94(0.11) & 0.01(0.02) & 1.04(1.19)  \\
MIP           & 8.87(16.48)  & -\,-         & -\,-         & 0.22(0.11) & 0.97(0.08) & 0.16(0.10) & -\,-            \\
Heaping ZIP   & 3.66(0.50)   & -\,-         & -\,-         & 0.55(0.02) & -\,-         & -\,-         & -\,-            \\
NB-LASSO      & 6.28(0.37)   & 0.06(0.00) & -\,-         & -\,-         & -\,-         & -\,-         & 11.38(3.71) \\
Poisson-LASSO & 4.08(0.81)   & 0.65(0.18) & -\,-         & -\,-         & -\,-         & -\,-         & -\,-            \\
NB            & 3.77(0.51)   & -\,-         & -\,-         & -\,-         & -\,-         & -\,-         & 3.58(0.70)  \\
\hline
\end{tabular}}
\end{table}

\begin{table}[htbp]
\caption{   Simulation results under scenario 8 with different sample sizes. In each cell, mean(SD) based on 100 replicates.}
\setlength{\tabcolsep}{1.5pt}
\renewcommand\arraystretch{1}
\resizebox{\textwidth}{!}{
\begin{tabular}{lccccccc}
\hline
Approach        & \multicolumn{1}{c}{RSSE:C}  & \multicolumn{1}{c}{TPR:C} & \multicolumn{1}{c}{FPR:C} & \multicolumn{1}{c}{ RSSE:I} & \multicolumn{1}{c}{TPR:I} & \multicolumn{1}{c}{FPR:I}  & \multicolumn{1}{c}{AE:D} \\ \hline
              & \multicolumn{7}{c}{$n =$ 500}                                                                 \\

proposed      & 1.40(0.37) & 0.83(0.11) & 0.12(0.16) & 0.12(0.11) & 1.00(0.00) & 0.01(0.02) & 0.73(0.95)     \\
MIP           & 1.98(0.77) & -\,-         & -\,-         & 0.15(0.07) & 1.00(0.00) & 0.10(0.07) & -\,-             \\
Heaping ZIP   & 1.66(0.33) & -\,-         & -\,-         & 1.75(0.00) & -\,-         & -\,-         & -\,-             \\
NB-LASSO      & 5.44(0.22) & 0.09(0.00) & 0.00(0.00) & -\,-         & -\,-         & -\,-         & 22.32(6.68) \\
Poisson-LASSO & 3.77(0.84) & 0.53(0.22) & 0.13(0.17) & -\,-         & -\,-         & -\,-         & -\,-             \\
NB            & 1.54(0.34) & -\,-         & -\,-         & -\,-         & -\,-         & -\,-         & 2.41(0.94)   \\
\hline

\end{tabular}}
\end{table}

\begin{table}[htbp]
\caption{   Simulation results under scenario 9 with different sample sizes. In each cell, mean(SD) based on 100 replicates.}
\setlength{\tabcolsep}{1.5pt}
\renewcommand\arraystretch{1}
\resizebox{\textwidth}{!}{
\begin{tabular}{lccccccc}
\hline
Approach        & \multicolumn{1}{c}{RSSE:C}  & \multicolumn{1}{c}{TPR:C} & \multicolumn{1}{c}{FPR:C} & \multicolumn{1}{c}{ RSSE:I} & \multicolumn{1}{c}{TPR:I} & \multicolumn{1}{c}{FPR:I}  & \multicolumn{1}{c}{AE:D} \\ \hline
              & \multicolumn{7}{c}{$n =$ 500}                                                                 \\

proposed      & 1.11(0.70)  & 0.86(0.16) & 0.23(0.22) & 0.07(0.07) & 0.96(0.10) & 0.00(0.02) & 0.36(0.71)  \\
MIP           & 1.44(4.58)  & -\,-         & -\,-         & 0.06(0.06) & 0.98(0.05) & 0.02(0.07) & -\,-           \\
Heaping ZIP   & 2.99(0.50)  & -\,-         & -\,-         & 0.54(0.01) & -\,-         & -\,-         & -\,-           \\
NB-LASSO      & 5.53(0.35)  & 0.11(0.09) & 0.01(0.06) & -\,-         & -\,-         & -\,-         & 8.44(2.31) \\
Poisson-LASSO & 4.00(0.73)  & 0.60(0.17) & 0.12(0.17) & -\,-         & -\,-         & -\,-         & -\,-           \\
NB            & 3.39(0.47)  & -\,-         & -\,-         & -\,-         & -\,-         & -\,-         & 3.43(0.79)  \\
\hline

\end{tabular}}
\end{table}

\begin{table}[htbp]
\caption{   Simulation results under scenario 10 with different sample sizes. In each cell, mean(SD) based on 100 replicates.}
\setlength{\tabcolsep}{1.5pt}
\renewcommand\arraystretch{1}
\resizebox{\textwidth}{!}{
\begin{tabular}{lccccccc}
\hline
Approach        & \multicolumn{1}{c}{RSSE:C}  & \multicolumn{1}{c}{TPR:C} & \multicolumn{1}{c}{FPR:C} & \multicolumn{1}{c}{ RSSE:I} & \multicolumn{1}{c}{TPR:I} & \multicolumn{1}{c}{FPR:I}  & \multicolumn{1}{c}{AE:D} \\ \hline
              & \multicolumn{7}{c}{$n =$ 500}                                                                 \\

proposed      & 1.12(0.61)  & 0.83(0.17) & 0.15(0.19) & 0.06(0.04) & 1.00(0.00) & 0.00(0.01) &  0.18(0.44) \\
MIP           & 1.38(3.98)  & -\,-         & -\,-         & 0.06(0.05) & 1.00(0.00) & 0.02(0.07) & -\,-           \\
Heaping ZIP   & 3.31(0.43)  & -\,-         & -\,-         & 0.75(0.01) & -\,-         & -\,-         & -\,-           \\
NB-LASSO      & 8.18(24.15) & 0.12(0.05) & 0.00(0.00) & -\,-         & -\,-         & -\,-         & 5.53(2.04) \\
Poisson-LASSO & 4.19(0.75)  & 0.56(0.19) & 0.12(0.19) & -\,-         & -\,-         & -\,-         & -\,-           \\
NB            & 3.54(0.33)  & -\,-         & -\,-         & -\,-         & -\,-         & -\,-         &  2.22(0.44) \\
\hline

\end{tabular}}
\end{table}

\begin{table}[htbp]
\caption{   Simulation results under scenario 11 with different sample sizes. In each cell, mean(SD) based on 100 replicates.}
\setlength{\tabcolsep}{1.5pt}
\renewcommand\arraystretch{1}
\resizebox{\textwidth}{!}{
\begin{tabular}{lccccccc}
\hline
Approach        & \multicolumn{1}{c}{RSSE:C}  & \multicolumn{1}{c}{TPR:C} & \multicolumn{1}{c}{FPR:C} & \multicolumn{1}{c}{ RSSE:I} & \multicolumn{1}{c}{TPR:I} & \multicolumn{1}{c}{FPR:I}  & \multicolumn{1}{c}{AE:D} \\ \hline
              & \multicolumn{7}{c}{$n =$ 500}                                                                 \\
proposed      & 2.92(1.06) & 0.55(0.27) & 0.21(0.35) & 0.22(0.17) & 1.00(0.00) & 0.00(0.02) & 3.81(5.51) \\
MIP           & 2.94(4.51) & -\,-         & -\,-         & 0.07(0.04) & 1.00(0.00) & 0.01(0.05) & -\,-            \\
Heaping ZIP   & 4.65(0.42) & -\,-         & -\,-         & 0.92(0.04) & -\,-         & -\,-         & -\,-            \\
NB-LASSO      & 5.45(0.11) & 0.10(0.03) & 0.00(0.02) & -\,-         & -\,-         & -\,-         & 5.87(1.44)  \\
Poisson-LASSO & 5.08(0.56) & 0.28(0.22) & 0.06(0.14) & -\,-         & -\,-         & -\,-         & -\,-            \\
NB            & 4.47(0.29) & -\,-         & -\,-         & -\,-         & -\,-         & -\,-         & 4.22(0.70)  \\
\hline

\end{tabular}}
\end{table}

\begin{table}[htbp]
\caption{   Simulation results under scenario 12 with different sample sizes. In each cell, mean(SD) based on 100 replicates.}
\setlength{\tabcolsep}{1.5pt}
\renewcommand\arraystretch{1}
\resizebox{\textwidth}{!}{
\begin{tabular}{lccccccc}
\hline
Approach        & \multicolumn{1}{c}{RSSE:C}  & \multicolumn{1}{c}{TPR:C} & \multicolumn{1}{c}{FPR:C} & \multicolumn{1}{c}{ RSSE:I} & \multicolumn{1}{c}{TPR:I} & \multicolumn{1}{c}{FPR:I}  & \multicolumn{1}{c}{AE:D} \\ \hline
              & \multicolumn{7}{c}{$n =$ 500}                                                                 \\
proposed      & 0.70(0.66)    & 0.96(0.08) & 0.29(0.22) & 0.06(0.06) & 0.97(0.09) & 0.00(0.01) & 0.24(0.88)   \\
MIP           & 1.54(9.19)    & -\,-         & -\,-         & 0.05(0.03) & 0.99(0.03) & 0.01(0.05) & -\,-            \\
Heaping ZIP   & 3.03(0.61)    & -\,-         & -\,-         & 0.54(0.01) & -\,-         & -\,-         & -\,-            \\
NB-LASSO      & 6.46(0.25)    & 0.09(0.00) & 0.00(0.00) & -\,-         & -\,-         & -\,-         & 14.31(3.56)   \\
Poisson-LASSO & 3.86(0.81)    & 0.81(0.13) & 0.29(0.23) & -\,-         & -\,-         & -\,-         & -\,-            \\
NB            & 3.69(0.51)    & -\,-         & -\,-         & -\,-         & -\,-         & -\,-         & 3.72(0.80)   \\
\hline

\end{tabular}}
\end{table}

\begin{table}[htbp]
\caption{   Simulation results under scenario 13 with different sample sizes. In each cell, mean(SD) based on 100 replicates.}
\setlength{\tabcolsep}{1.5pt}
\renewcommand\arraystretch{1}
\resizebox{\textwidth}{!}{
\begin{tabular}{lccccccc}
\hline
Approach        & \multicolumn{1}{c}{RSSE:C}  & \multicolumn{1}{c}{TPR:C} & \multicolumn{1}{c}{FPR:C} & \multicolumn{1}{c}{ RSSE:I} & \multicolumn{1}{c}{TPR:I} & \multicolumn{1}{c}{FPR:I}  & \multicolumn{1}{c}{AE:D} \\ \hline
              & \multicolumn{7}{c}{$n =$ 500}                                                                 \\
proposed      & 0.60(0.35)   & 0.99(0.04) & -\,-         & 0.05(0.03) & 0.98(0.07) & 0.00(0.02) & 0.07(0.36)    \\
MIP           & 2.17(9.53)   & -\,-         & -\,-         & 0.05(0.04) & 0.99(0.05) & 0.01(0.06) & -\,-            \\
Heaping ZIP   & 3.01(0.60)   & -\,-         & -\,-         & 0.54(0.01) & -\,-         & -\,-         & -\,-            \\
NB-LASSO      & 6.21(0.54)   & 0.08(0.10) & -\,-         & -\,-         & -\,-         & -\,-         & 10.28(3.15) \\
Poisson-LASSO & 4.01(0.74)   & 0.69(0.16) & -\,-         & -\,-         & -\,-         & -\,-         & -\,-            \\
NB            & 3.56(0.50)   & -\,-         & -\,-         & -\,-         & -\,-         & -\,-         & 3.62(0.70) \\
\hline

\end{tabular}}
\end{table}

\begin{table}[htbp]
\caption{   Simulation results under scenario 14 with different sample sizes. In each cell, mean(SD) based on 100 replicates.}
\setlength{\tabcolsep}{1.5pt}
\renewcommand\arraystretch{1}
\resizebox{\textwidth}{!}{
\begin{tabular}{lccccccc}
\hline
Approach        & \multicolumn{1}{c}{RSSE:C}  & \multicolumn{1}{c}{TPR:C} & \multicolumn{1}{c}{FPR:C} & \multicolumn{1}{c}{ RSSE:I} & \multicolumn{1}{c}{TPR:I} & \multicolumn{1}{c}{FPR:I}  & \multicolumn{1}{c}{AE:D} \\ \hline
              & \multicolumn{7}{c}{$n =$ 500}                                                                 \\
proposed      & 0.52(0.20) & 0.99(0.04) & 0.13(0.17) & 0.04(0.03) & 1.00(0.00) & 0.01(0.02) & 0.00(0.01)     \\
MIP           & 0.59(0.20) & -\,-         & -\,-         & 0.04(0.03) & 1.00(0.00) & 0.01(0.02) & -\,-             \\
Heaping ZIP   & 0.58(0.17) & -\,-         & -\,-         & 1.74(0.00) & -\,-         & -\,-         & -\,-             \\
NB-LASSO      & 5.45(0.20) & 0.09(0.00) & 0.00(0.00) & -\,-         & -\,-         & -\,-         & 18.78(4.81) \\
Poisson-LASSO & 3.66(0.81) & 0.57(0.19) & 0.10(0.18) & -\,-         & -\,-         & -\,-         & -\,-             \\
NB            & 1.16(0.25) & -\,-         & -\,-         & -\,-         & -\,-         & -\,-         & 1.98(0.45)  \\
\hline

\end{tabular}}
\end{table}

\begin{table}[htbp]
\caption{   Simulation results under scenario 15 with different sample sizes. In each cell, mean(SD) based on 100 replicates.}
\setlength{\tabcolsep}{1.5pt}
\renewcommand\arraystretch{1}
\resizebox{\textwidth}{!}{
\begin{tabular}{lccccccc}
\hline
Approach        & \multicolumn{1}{c}{RSSE:C}  & \multicolumn{1}{c}{TPR:C} & \multicolumn{1}{c}{FPR:C} & \multicolumn{1}{c}{ RSSE:I} & \multicolumn{1}{c}{TPR:I} & \multicolumn{1}{c}{FPR:I}  & \multicolumn{1}{c}{AE:D} \\ \hline
              & \multicolumn{7}{c}{$n =$ 500}                                                                 \\
proposed      & 0.28(0.10)   & 1.00(0.00) & 0.09(0.12) & 0.01(0.01) & -\,-         & 0.05(0.02) & 0.00(0.00)   \\
MIP           & 0.74(0.54)   & -\,-         & -\,-         & 0.00(0.01) & -\,-         & 0.03(0.03) & -\,-           \\
Heaping ZIP   & 0.34(0.09)   & -\,-         & -\,-         & 0.00(0.00) & -\,-         & -\,-         & -\,-           \\
NB-LASSO      & 5.86(19.51)  & 0.48(0.42) & 0.01(0.04) & -\,-         & -\,-         & -\,-         & 5.71(5.21) \\
Poisson-LASSO & 2.71(0.55)   & 0.81(0.11) & 0.06(0.11) & -\,-         & -\,-         & -\,-         & -\,-           \\
NB            & 0.32(0.09)   & -\,-         & -\,-         & -\,-         & -\,-         & -\,-         & 0.00(0.00)   \\
\hline

\end{tabular}}
\end{table}